\documentclass[journal]{IEEEtran}
\ifCLASSINFOpdf
\else
\fi
\usepackage[linesnumbered,ruled,vlined]{algorithm2e}
\usepackage{url}
\usepackage{array}
\usepackage{graphicx}
\usepackage{etoolbox}
\usepackage{hyperref}
\usepackage{ctable}
\usepackage{multirow}
\usepackage{booktabs}
\usepackage{threeparttable}
\usepackage{amsthm}
\usepackage{amssymb}
\usepackage{mathtools}
\usepackage{cleveref}
\usepackage[noadjust]{cite}
\usepackage{lettrine}
\usepackage{verbatim}
\usepackage{algpseudocode}
\usepackage{hyperref}
\usepackage{pifont}
\makeatletter
\long\def\@makecaption#1#2{\ifx\@captype\@IEEEtablestring%
	\footnotesize\begin{center}{\normalfont\footnotesize #1}\\
		{\normalfont\footnotesize\scshape #2}\end{center}%
	\@IEEEtablecaptionsepspace
	\else
	\@IEEEfigurecaptionsepspace
	\setbox\@tempboxa\hbox{\normalfont\footnotesize {#1.}~~ #2}%
	\ifdim \wd\@tempboxa >\hsize%
	\setbox\@tempboxa\hbox{\normalfont\footnotesize {#1.}~~ }%
	\parbox[t]{\hsize}{\normalfont\footnotesize \noindent\unhbox\@tempboxa#2}%
	\else
	\hbox to\hsize{\normalfont\footnotesize\hfil\box\@tempboxa\hfil}\fi\fi}
\makeatother
\usepackage{dblfloatfix}
\makeatletter
\newcommand{\algorithmfootnote}[2][\footnotesize]{%
	\let\old@algocf@finish\@algocf@finish
	\def\@algocf@finish{\old@algocf@finish
		\leavevmode\rlap{\begin{minipage}{\linewidth}
				#1#2
		\end{minipage}}%
	}%
}
\makeatother

\undef{\proof}
\newtheorem*{proof}{\textbf{Proof}}

\undef{\corollary}
\newtheorem{corollary}{\textbf{Corollary}}
\usepackage{todonotes}
\usepackage{gensymb}

\usepackage{multicol}

\renewcommand\IEEEkeywordsname{Keywords}
\usepackage{flushend}
\usepackage{color, colortbl}
\definecolor{Gray}{gray}{0.9}
\begin{document}
%
\title{Dilution with Digital Microfluidic Biochips: How Unbalanced Splits Corrupt Target-Concentration}
%
%
%
\author{Sudip Poddar, Robert Wille, Hafizur Rahaman, and Bhargab B. Bhattacharya

\thanks{S. Poddar is with the Advanced Computing and
	Microelectronics Unit, Indian Statistical Institute, Kolkata, India 700108. Email: sudippoddar2006@gmail.com}
\thanks{Bhargab B. Bhattacharya is with the Department of Computer Science \& Engineering, Indian Institute of Technology Kharagpur, India 721 302; this work was done while he had been with  Indian Statistical Institute, Kolkata, India 700108. Email:  bhargab.bhatta@gmail.com.}
\thanks{H. Rahaman is with the School of VLSI Technology, Indian Institute
	of Engineering Science and Technology, Shibpur, India 711103. E-mail:
	hafizur@vlsi.iiests.ac.in.}
\thanks{R. Wille is with the Institute for Integrated Circuits, Johannes Kepler University Linz, Austria. E-mail: robert.wille@jku.at.}
 }

%
%

\markboth{Dilution with Digital Microfluidic Biochips: How Unbalanced Splits Corrupt Target-Concentration}{Poddar
	\MakeLowercase{\textit{et al.}}}
%



\maketitle




%
\IEEEpeerreviewmaketitle
\begin{abstract}
Sample preparation is an indispensable component of almost all biochemical protocols, and it involves, among others, making dilutions and mixtures of fluids in certain ratios. Recent microfluidic technologies offer suitable platforms for automating dilutions on-chip, and typically on a digital microfluidic biochip~(DMFB), a sequence of $(1:1)$ mix-split operations is performed on fluid droplets to achieve the target concentration factor $(CF)$ of a sample. An $(1:1)$ mixing model ideally comprises mixing of two unit-volume droplets followed by a (balanced) splitting into two unit-volume daughter-droplets. However, a major source of error in fluidic operations is due to unbalanced splitting, where two unequal-volume droplets are produced following a split. Such volumetric split-errors occurring in different mix-split steps of the reaction path often cause a significant drift in the target-{\em CF} of the sample, the precision of which cannot be compromised in life-critical assays. In order to circumvent this problem, several error-recovery or error-tolerant techniques have been proposed recently for DMFBs. Unfortunately, the impact of such fluidic errors on a target-{\em CF} and the dynamics of their behavior have
not yet been rigorously analyzed. In this work, we investigate the effect of multiple volumetric split-errors on various target-{\em CF}s during sample preparation. We also perform a detailed analysis of the worst-case scenario, i.e., the condition when the error in a target-{\em CF} is maximized. This analysis may lead to the development of new techniques for error-tolerant sample preparation with DMFBs without using any sensing operation.

\end{abstract}

%
%
%
%
%
%
%
%
\noindent
\section{Introduction}
A digital microfluidic biochip (DMFB) is capable of executing multiple tasks of biochemical laboratory protocols in an efficient manner. DMFBs support
droplet-based operations on a single chip with high sensitivity
and reconfigurability. Discrete
volume (nanoliter/picoliter) droplets are manipulated on DMFBs through
electrical actuation on an electrode array~\cite{EWOD}. Various fluid-handling operations such as dispensing, transport, mixing, split, dilution can be performed on these tiny chips with higher speed and reliability. Due to their versatile properties,
these programmable chips are used in many applications such as
{\em in-vitro} diagnostics (point-of-care, self-testing),
drug discovery (high-throughput screening), biotechnology (process monitoring, process development), ecology (agriculture, environment, homeland security), and sample preparation~\cite{b19,B403341H,5487469,Alistar:2015,BS}.

Sample preparation imparts significant impact on accuracy,
assay-completion time and cost, and 
plays a pivotal role in biomedical engineering and life science~\cite{Dynamic}.
It involves dilution or mixture preparation and comprises a sequence of mixing steps necessary to produce
	a mixture of input reagents having a desired ratio of the constituents. Note that sample collection, transportation, and preparation consume up to 90\% cost and 95\% of time~\cite{Poddar_2016}.
In last few years, a large
number of sample-preparation algorithms had been developed for reducing assay-completion time
and cost~\cite{BS,DMRW,REMIA,WARA,MTC,FloSPA,RSM,stor_aware_2018}, based on the (1:1) mixing model and similar.
In the conventional (1:1) mixing model, two unit-volume of droplets are mixed together and split into two equal-sized daughter droplets following the mixing
operation. These algorithms output a particular sequence of mix-split operations~(represented as a sequencing graph) in order to dilute the target-droplet to a desired target-concentration. For the convenience of dilution
algorithms, a concentration factor ({\em CF}) is
approximated with a binary fraction, which is
reachable and satisfies a user-defined error-tolerance limit. The detailed
description of sample preparation can be found elsewhere~\cite{error_oblivious}.

Although droplet-based microfluidic biochips enable the integration of fluid-handling operations and outcome sensing on a single biochip, errors are likely to occur during fluidic
operations due to various permanent faults (e.g., dielectric breakdown or charge
trapping), or transient
faults (e.g., unbalanced split due to imperfect actuation). For
example, two daughter-droplets may be of different volume after split-operation
while executing mix-split steps on a DMFB platform. The
unequal-volume droplets produced after an erroneous mix-split step, when used later will negatively impact the correctness of the desired target-{\em CF}. Therefore, unbalanced-split errors pose a significant threat to sample preparation. Hence, from the viewpoint of error-management, it is essential to introduce some error-management scheme to handle such faults during sample preparation.

 \begin{figure*}[!ht]
 	\centering
 	\includegraphics[width=10.8cm, height = 3.5cm]{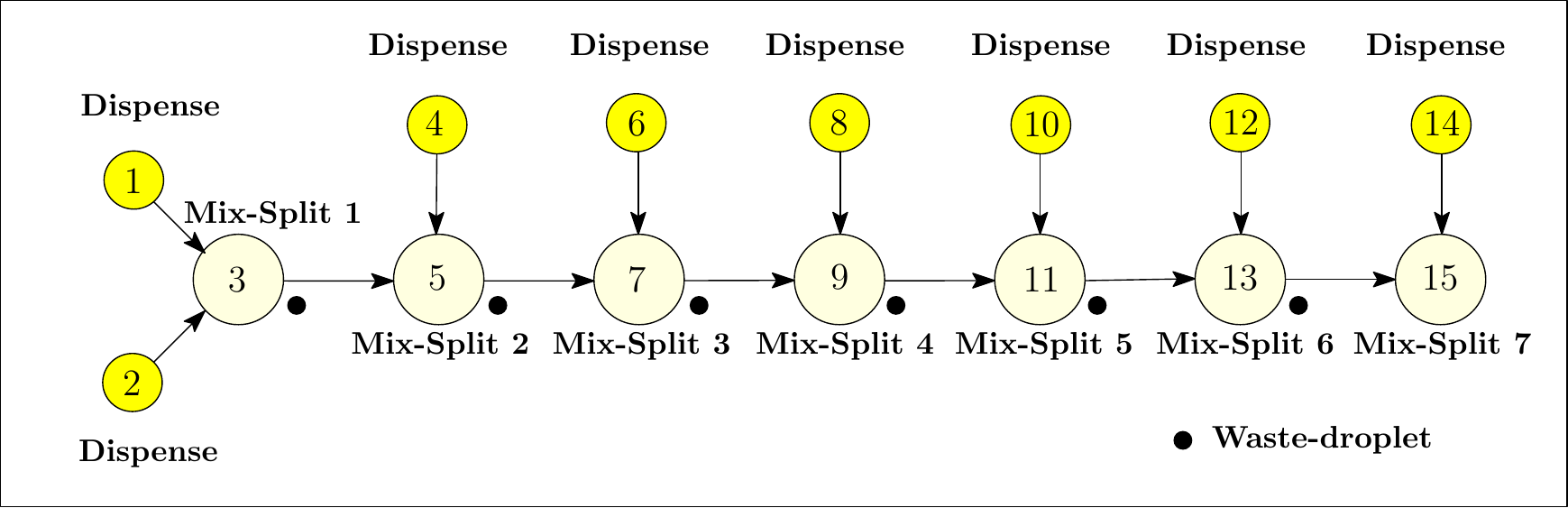}
 	\caption{Initial sequencing graph.}
 	\label{fig:seq_graph}
 \end{figure*}

In this paper, we focus especially on volumetric split-errors and investigate their effects on
the target-{\em CF} during sample preparation.
Split-errors may unexpectedly occur in any mix-split
step of the mixing-path during sample preparation, thus affecting 
the concentration factor~({\em CF}) of the target-droplet~\cite{Poddar_2016}.
Moreover, due to the unpredictable characteristics 
of fluidic-droplets, a daughter droplet of larger or smaller size 
may be used in the mixing path following an erroneous split operation{\footnote{depending on the  
selection of the erroneous droplet (larger or smaller volume) to be used 
in a subsequent step.}} on the mixing-path. Although a number of cyber-physical based approaches  have been proposed for error-recovery~\cite{control_path,cyber_physical,Luo_dictionary,LuoTCAD14Uncertainity,Dynamic}, they do not provide
any guarantee on the number of rollback iterations that are needed to rectify the error. Thus, most of the prior approaches to error-recovery 
in biochips are non-deterministic in nature. On the other hand, the approach
proposed in 
\cite{Poddar_2016} performs error-correction in a deterministic sense;
however, it assumes only the presence of single split-errors while
classifying them
as being {\em critical} or {\em non-critical}. A split-error occurring
at a particular step is called {\em critical} ({\em non-critical}), if 
a single split-error when inserted at the corresponding step, causes
the target-{\em CF} to exceed (bound within) the allowable error-tolerance
range. This approach does not consider the possibility of multiple split-errors during classification. Furthermore, in a cyber-physical
settings, 
it requires some additional
time for sensing the occurrence of a {\em critical} error, if any,
at every such step. Hence, when the number of
{\em {\em critical}} errors becomes large, sensing time may outweigh the gain obtained in
roll-forwarding assay-time, and as a result, we may need a longer
overall execution time compared to that of the proposed method.

In this paper, we present a thorough analysis of the impact of multiple
split-errors on a given target-{\em CF}. 
Based on these observations, methods for sample preparation that can deal with split errors even without any sensors and/or rollback
can be derived. In fact, the findings discussed in this paper yield a 
method (described in~\cite{error_oblivious}) that produces a target-{\em CF} within the allowable
error-tolerance limit without using any sensor.

The remainder of the paper is organized as follows. Section~\ref{sec:prior_art} introduces the basic principle of earlier error-recovery approaches. We describe the effect of one or more
volumetric split-errors on the target-{\em CF}, in Section~\ref{sec:error_effect}. Section~\ref{sec:max_error} presents the worst-case scenario, i.e., when {\em CF}-error in the target-droplet becomes maximum.
A justification behind the maximum {\em CF}-error is then reported in
Section~\ref{sec:max_error_justifiaction}. Finally, we draw our 
conclusions in Section~\ref{sec:conclusion}.

\section{Error-recovery approaches: prior art}
\label{sec:prior_art}
Earlier approaches perform error-recovery operations
by repeating the concerned operations of the bioassay~\cite{Mein2000} for producing the target concentration factor within the allowable error-range. For example, all 
mix-split operations and dispensing operations of the initial sequencing graph~(shown in Fig.~\ref{fig:seq_graph}) were re-executed when an error is detected 
at the end (after execution of the bio-assay). However, the repetition of
such experiments leads to wastage of precious
reagents and hard-to-obtain samples, and results in longer assay completion-time.

\subsection{Cyber-physical technique for error-recovery}
\label{sec:des_cyber_phycal}
\begin{figure}[t]
	\centering
	\includegraphics[width=10.0cm, height = 4.0cm]{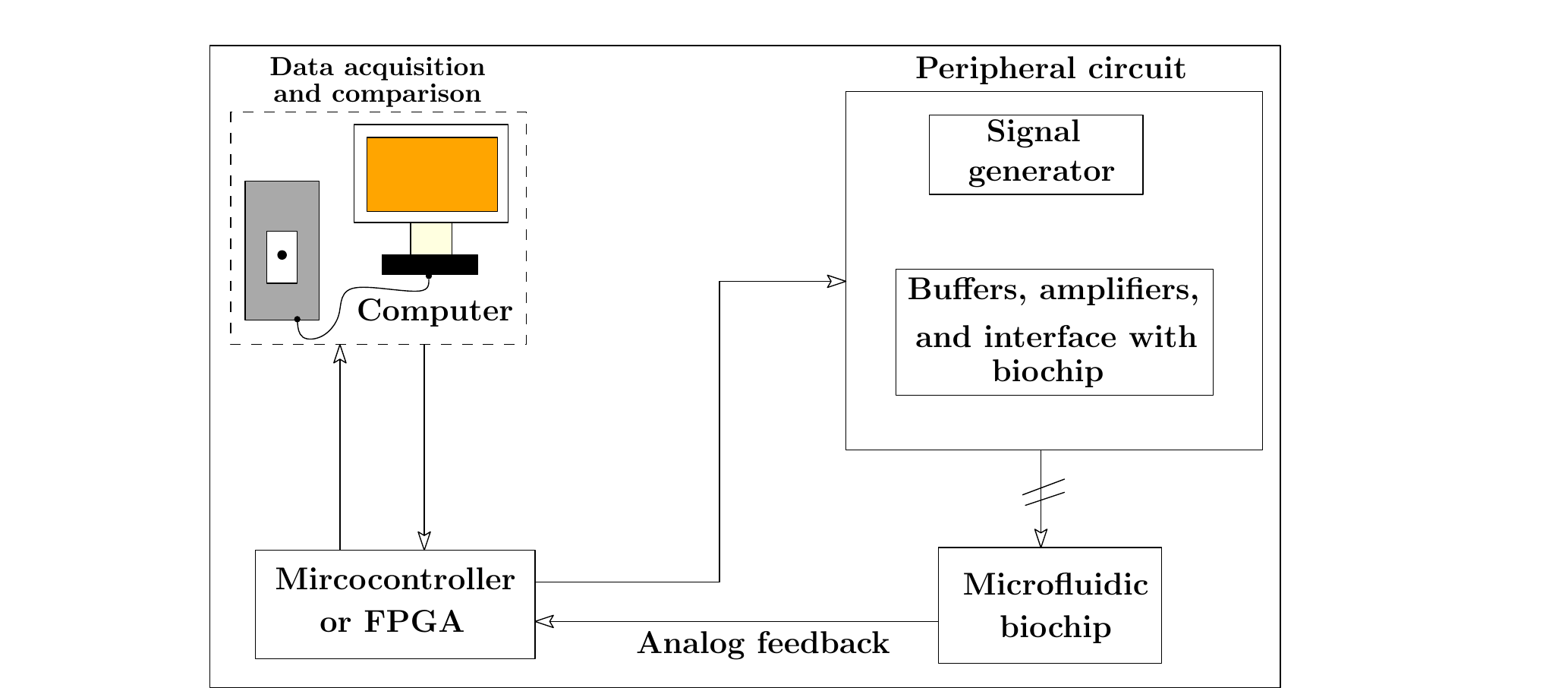}
	\caption{Schematic of a cyberphysical error-recovery system.}
	\label{fig:cyber_system}
\end{figure}

In order to avoid such repetitive execution of on-chip biochemical experiments, recently, cyber-physical {DMFB}s were proposed for
obtaining the desired outcome~\cite{cyber_physical}. A diagram of a cyber-physical
biochip is shown in Fig.~\ref{fig:cyber_system} for demonstration purpose.
It consists of the following components: a computer, a single-board
microcontroller or an FPGA, a peripheral circuit, and the concerned biochip. Two
interfaces are required for establishing the connection 
between control software and hardware of the microfluidic system. The first interface is required for converting the output signal of the
sensor to an input signal that feeds the control software installed on the
computer. The second interface transforms the output of the
control software into a sequence of voltage-actuation maps that activate the
electrodes of the biochip. The error-recovery operation is executed by
the control software running in the back-end.

\begin{figure*}[!ht]
	\centering
	\includegraphics[width=10.8cm, height = 5.2cm]{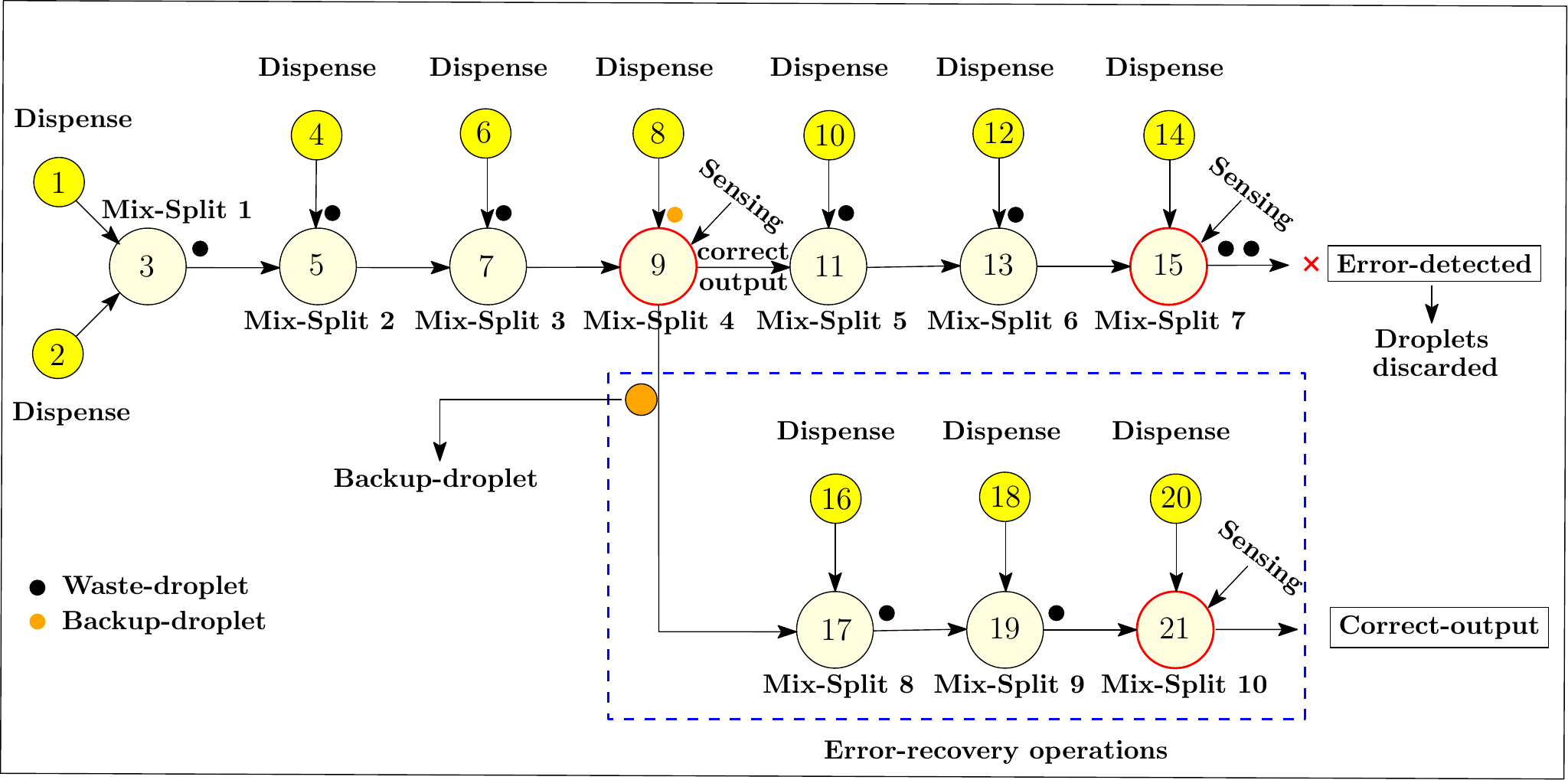}
	\caption{Generation of a target-droplet by cyber-physical  error-recovery approaches.}
	\label{fig:cyber_methods}
\end{figure*}

\subsection{Compilation  for error-recovery}
Note that cyber-physical based {DMFB}s need to
constantly monitor the output of the intermediate mix-split operations
at designated checkpoints using on-chip sensors (integrated with the biochip). The original actuation sequences are interrupted when an error is detected during the execution of a bioassay. 
At the same time, the recovery actions, e.g., the re-execution of corresponding
dispensing and mixing operations is initiated to remedy the error. 
However, the error-recovery operations will have to be translated into electrode-actuation sequences in real-time.

The compilation of error-recovery actions can 
either be performed before actual execution of the bio-assay or during the
execution of the bio-assay. So, depending on the compilation-time of
operations, error-recovery approaches can be divided into two
categories:
i) offline (at design time), and ii) online (at run time). 

In the offline
approach, all possible errors of interest that might occur (under the assumed 
model) during the 
execution of a bio-assay are identified,  and  compilation is performed to pre-compute
and store the corresponding error-recovery actuation sequences. They will
provide an alternative schedule, which is stored in the memory. When 
an error is detected during actual execution of the bio-assay, the 
cyber-physical biochip executes the error-recovery actions by loading
the corresponding schedules from the memory. However, this approach
can be used to rectify only a limited number of errors ($\leq$ 2) since 
a very large-size controller memory will be required to store the recovery sequences for all possible
consequences of errors~\cite{Luo_dictionary}.

On the other hand, in the 
online approach, appropriate actions are carried out depending on the
feedback given by the sensor. Compilation of error-recovery actions 
into electrode-actuation sequences is performed only at run-time.

\subsection{Working principle of cyber-physical based DMFBs}
In spite of the above difference, cyber-physical {DMFB}s perform error-recovery
operations as follows. During
actual execution of the bio-assay, a biochip receives control signals
from the software running on the computer system. At the same time, the 
sensing
system of the biochip sends a feedback signal to the software by processing 
it using field-programmable gate array~(FPGA), or ASIC chips. If an error is detected by a sensor, the
control software immediately discards the erroneous-droplet for preventing
error-propagation, and performs the necessary error-recovery operations~(i.e., corresponding actuation sequences are determined
online/offline) for generating the correct output.

\begin{figure}[t]
	\centering
	\fbox{\includegraphics[width=7.2cm, height = 7.20cm]{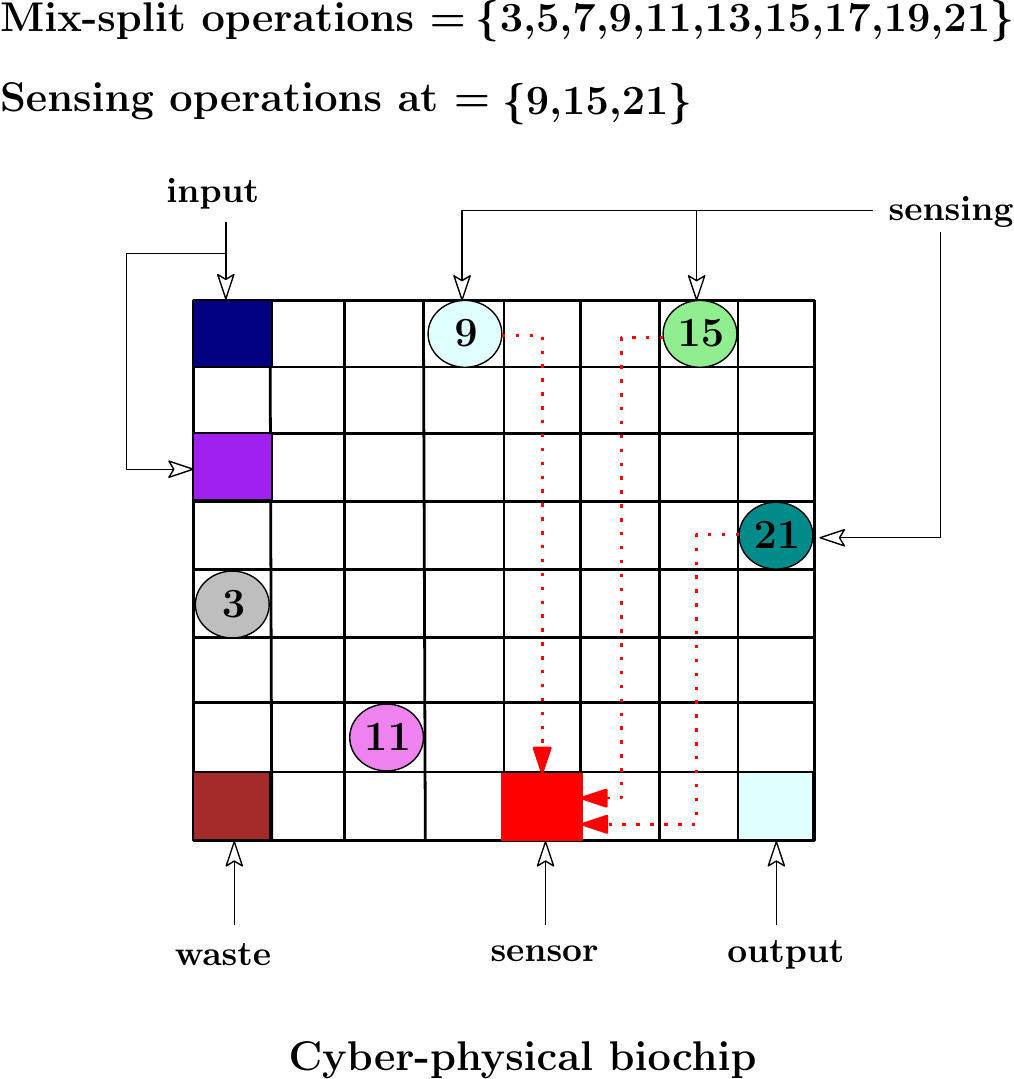}}
\caption{Routing of droplets for sensing operation in a cyber-physical biochip.}
\label{fig:cyber_biochips}
\end{figure}

In order to produce the correct output, the outcome of the 
intermediate mix-split operations are verified
using on-chip sensors suitably placed at designated checkpoints. 
For example, in Fig.~\ref{fig:cyber_methods}, the outcomes of 
{\em Mix-split 4} and
{\em Mix-split 7} are checked by the sensor. When an error is
detected, a portion of the bio-assay is re-executed. For instance, the 
operations shown within the blue box in Fig.~\ref{fig:cyber_methods} are
re-executed when an error is detected at the last checkpoint. 
Note that the accuracy of a cyberphysical system also depends on the
sensitivity of sensors. Unfortunately, due to cost constraints, only a
limited number of sensors can be integrated into a
{DMFB}~\cite{cyber_physical}. Additionally, in order to check 
the status of intermediate droplets, they need to be 
routed to a designated sensor location on the chip. This may introduce a
significant latency to the overall assay-completion
time~(Fig.~\ref{fig:cyber_biochips}).  As a result, prior cyber-physical based 
error-recovery methods for sample preparation become expensive in terms of
assay-completion time and reagent cost.

\begin{figure*}[!ht]
	\centering
	\includegraphics[width=10.8cm, height = 3.5cm]{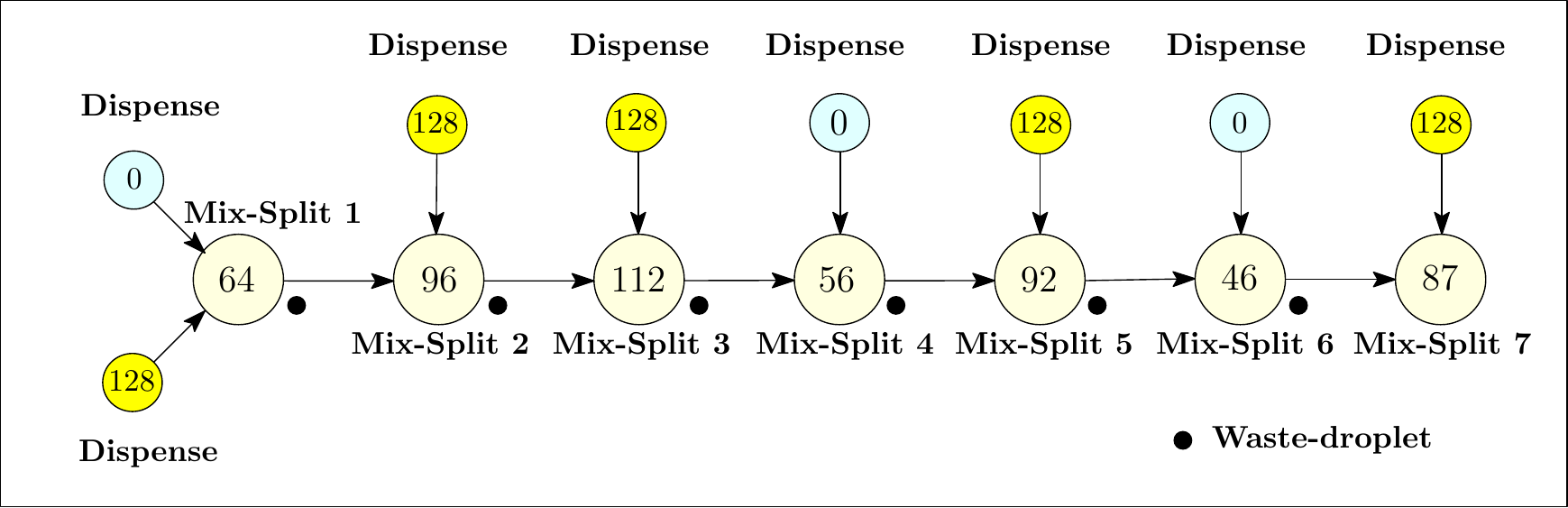}
	\caption{Sequence of mix-split operations for the target-{\em CF} = $\frac{87}{128}$.}
	\label{fig:example_cf}
\end{figure*} 

\begin{figure*}[!ht]
	\centering
	\includegraphics[width=10.8cm, height = 6.0cm]{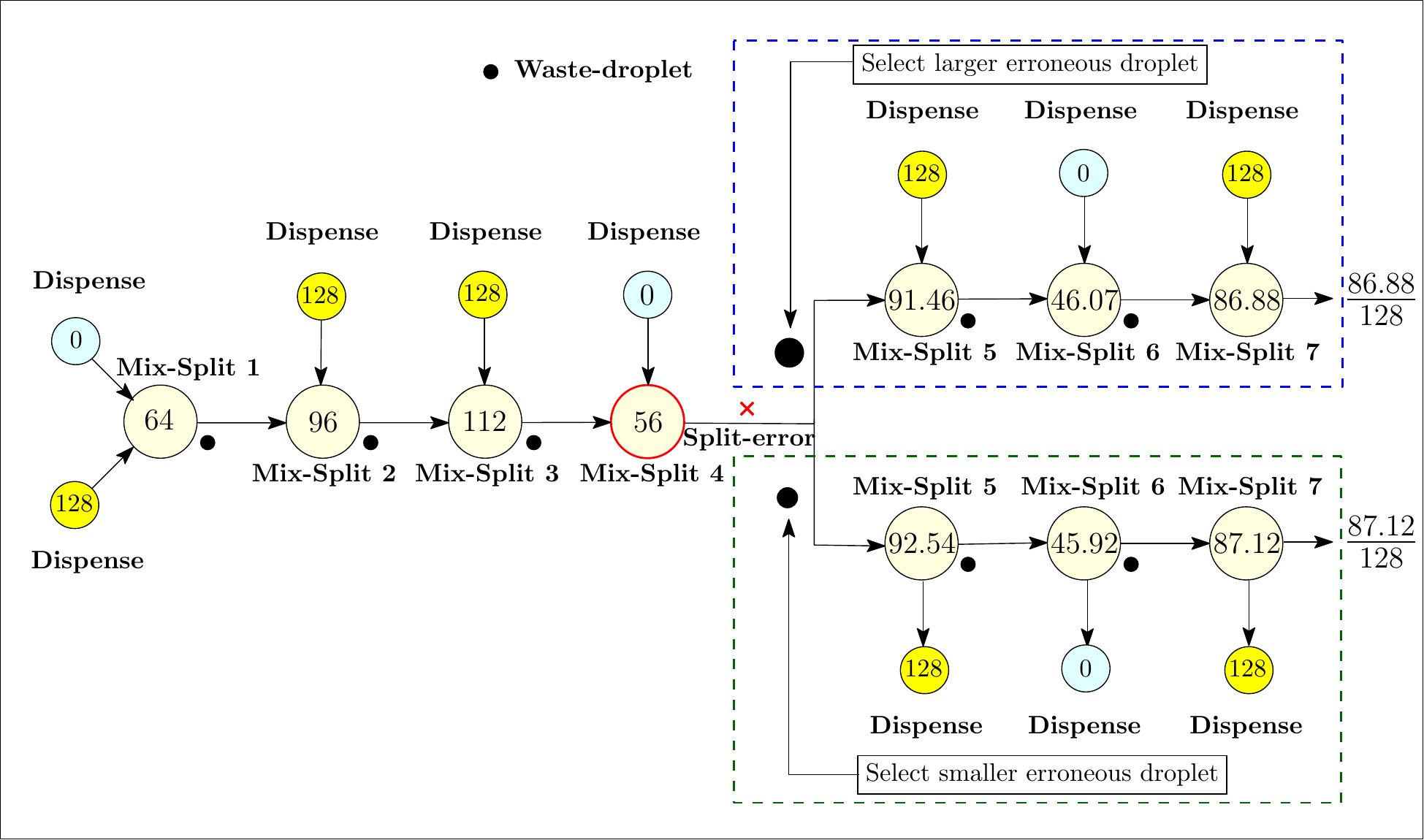}
	\caption{Effect of choosing larger-/smaller-volume erroneous droplet on the target-{\em CF} = $\frac{87}{128}$.}
	\label{fig:error_example_cf}
\end{figure*}

To summarize, cyber-physical error-recovery methods suffer from the following shortcomings:
\begin{itemize}
\item They are expensive in terms of
assay-completion time and reagent-cost. Hence, they are unsuitable for
field deployment 
and point-of-care testing in resource-constrained areas.

\item  Prior cyberphysical solutions  
fail to provide any guarantee on the number of rollback attempts, i.e., how
many iterations will be required to correct the error. Hence, 
error-recovery becomes non-deterministic.
\item  Each component used in the design of cyberphysical coupling may 
become a possible source of failure, which ultimately reduces the reliability of the biochip.
\end{itemize}


Now, we present below a detailed analysis of multiple volumetric split-errors and their effects on a target-{\em CF}. 

%
%
%





\section{Effect of split-errors\\on the target concentration} 
\label{sec:error_effect}
Generally, in the (1:1) mixing model (where two 1X-volume droplets are
used for mixing operation), two 1X-volume daughter-droplets are produced 
after each mix-split operation. One of them is used in the subsequent mix-split operation and another
one is discarded as waste droplet or stored for later use~\cite{BS}~(see Fig.~\ref{fig:seq_graph}). An erroneous mix-split operation may produce
two unequal-volume droplets. Unless an elaborate sensing
mechanism is used, it is not possible to predict which one of the resulting
droplets~(smaller/larger) is going to be used in the subsequent mix-split operation. Moreover, their effect on the target-{\em CF} becomes more
complex when multiple volumetric split-errors occur in the mix-split path.

\subsection{Single volumetric split-error}
\label{single_error}
In order to analyze the effect of single volumetric split-error on
the target-{\em CF}, we perform experiments with different erroneous 
 droplets and present 
the results in this section. 
We assume an example target-{\em CF} = $\frac{87}{128}$
of accuracy level = 7. The mix-split sequence that needs to be
performed using {\em twoWayMix} algorithm \cite{BS} for generating the target-{\em CF} is shown in
Fig.~\ref{fig:example_cf}.

Let us consider the scenario of injecting 7\% volumetric split-error
at Mix-Split Step 4. Two unequal-volume daughter droplets are produced
after this step when a split-error occurs. As stated earlier, it may not be 
possible to predict which droplet (smaller/larger) will be used for the 
mixing operation in the next step. The effect of the erroneous
droplet on the target-{\em CF} depends on the choice of the
daughter-droplet to be used next. For example, the effect of
3\% volumetric split-error (at Step 4) on the target-{\em CF} = $\frac{87}{128}$ is shown in Fig.~\ref{fig:error_example_cf}. 
The effect of two errors on the
target-{\em CF} (when the larger or smaller volume droplet is used 
at Mix-Split Step 4) is also shown in Fig.~\ref{fig:error_example_cf}. The blue (green) box represents the scenario when the next operation
is executed with the larger (smaller) erroneous droplet. It has been seen from
Fig.~\ref{fig:error_example_cf} that the {\em CF}-error in the target
increases when the smaller erroneous droplet is used in the mixing path
compared to the use of the larger one.

\setlength{\tabcolsep}{1.8pt}
\begin{table}[!h]
	\begin{center}
	\begin{threeparttable}
		\centering
		\caption{{Impact on target-{\em CF} = $\frac{87}{128}$ for different volumetric split-errors.}}
		\scriptsize
		\label{tab:error_per_3}
		\setlength{\extrarowheight}{.2em}
		\begin{tabular}{c|cc|c|c}
			\hline
			\multicolumn{5}{c}{Volumetric split-error = 3\%.}\\ 
			\hline
			\multirow{2}{*}{{Erroneous mix-split step}} & \multicolumn{2}{c|}{Selected-droplet}  &{Target-{\em CF}}$\times$128* & Within error-tolerance limit?\\ \cline{2-3}
			& Larger & Smaller &  & ({{\em CF}-error$\times$128$<$0.5}?)  \\ \hline
			1 &  \ding{51} & \ding{53} & 86.98& Yes\\
			1 &  \ding{53} & \ding{51} & 87.01& Yes\\
			2 &  \ding{51} & \ding{53} & 87.01& Yes\\
			2 &  \ding{53} & \ding{51} & 86.99& Yes\\
			3 &  \ding{51} & \ding{53} & 87.04& Yes\\
			3 &  \ding{53} & \ding{51} & 86.95& Yes\\
			4 &  \ding{51} & \ding{53} & 86.88& Yes\\
			4 &  \ding{53} & \ding{51} & 87.12& Yes\\
			5 &  \ding{51} & \ding{53} & 87.04& Yes\\
			5 &  \ding{53} & \ding{51} & 86.96& Yes\\
			6 &  \ding{51} & \ding{53} & 86.39& No\\
			6 &  \ding{53} & \ding{51} & 87.62& No\\
			
			\hline
		\end{tabular}
			\begin{tabular}{c|cc|c|c}
		\multicolumn{5}{c}{Volumetric split-error = 5\%.}\\ 
		\hline
		\multirow{2}{*}{{Erroneous mix-split step}} & \multicolumn{2}{c|}{Selected-droplet}  &{Target-{\em CF}}$\times$128 & Within error-tolerance limit?\\ \cline{2-3}
		& Larger & Smaller &  & ({{\em CF}-error$\times$128$<$0.5}?)  \\ \hline
		1 &  \ding{51} & \ding{53} & 86.98& Yes\\
		1 &  \ding{53} & \ding{51} & 87.02& Yes\\
		2 &  \ding{51} & \ding{53} & 87.01& Yes\\
		2 &  \ding{53} & \ding{51} & 86.98& Yes\\
		3 &  \ding{51} & \ding{53} & 87.08& Yes\\
		3 &  \ding{53} & \ding{51} & 86.92& Yes\\
		4 &  \ding{51} & \ding{53} & 86.81& Yes\\
		4 &  \ding{53} & \ding{51} & 87.19& Yes\\
		5 &  \ding{51} & \ding{53} & 87.06& Yes\\
		5 &  \ding{53} & \ding{51} & 86.94& Yes\\
		6 &  \ding{51} & \ding{53} & 86.00& No\\
		6 &  \ding{53} & \ding{51} & 88.05& No\\
		
	\end{tabular}
		\begin{tabular}{c|cc|c|c}
	\hline
	\multicolumn{5}{c}{Volumetric split-error = 7\%.}\\ 
	\hline
	\multirow{2}{*}{{Erroneous mix-split step}} & \multicolumn{2}{c|}{Selected-droplet}  &{Target-{\em CF}}$\times$128 & Within error-tolerance limit?\\ \cline{2-3}
	& Larger & Smaller &  & ({{\em CF}-error$\times$128$<$0.5}?)  \\ \hline
	1 &  \ding{51} & \ding{53} & 86.97& Yes\\
	1 &  \ding{53} & \ding{51} & 87.03& Yes\\
	2 &  \ding{51} & \ding{53} & 87.02& Yes\\
	2 &  \ding{53} & \ding{51} & 86.98& Yes\\
	3 &  \ding{51} & \ding{53} & 87.11& Yes\\
	3 &  \ding{53} & \ding{51} & 86.89& Yes\\
	4 &  \ding{51} & \ding{53} & 86.73& Yes\\
	4 &  \ding{53} & \ding{51} & 87.27& Yes\\
	5 &  \ding{51} & \ding{53} & 87.09& Yes\\
	5 &  \ding{53} & \ding{51} & 86.91& Yes\\
	6 &  \ding{51} & \ding{53} & 85.61& No\\
	6 &  \ding{53} & \ding{51} & 88.49& No\\
	
	\hline
\end{tabular}
		\begin{tablenotes}
			\item *: Results are shown up to two decimal places.
		\end{tablenotes}
	\end{threeparttable}
\end{center}
\end{table}

\begin{figure*}[!b]    
	\begin{center}
		\includegraphics[width=11.2cm, height = 3.5cm]{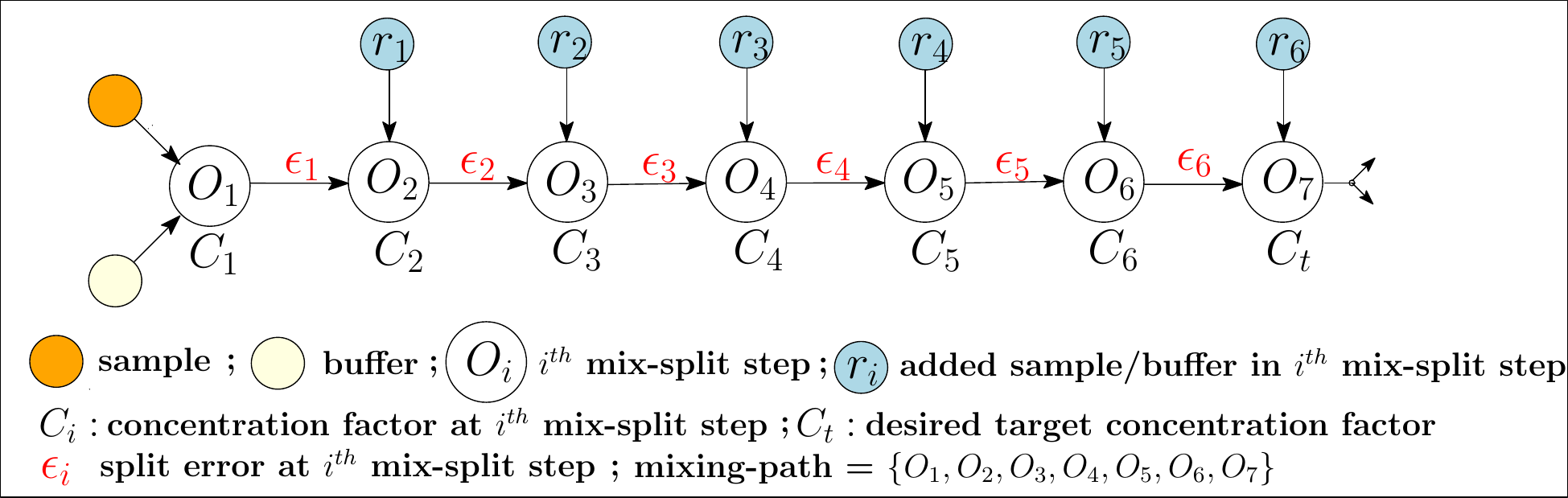}
		\caption{Mix-split operations for generating target-{\em CF}
			= $C_t$ with accuracy level $n$ = 7.}
		\label{fig:error}
	\end{center}
\end{figure*}

Similarly, we perform further experiments for finding the effect of
erroneous droplets on the target-{\em CF}. We report
the results for volumetric split-error 3\%, 5\% and 7\% 
occurring on the
mixing path,  in Table~\ref{tab:error_per_3}. We observe that the 
{\em CF}-error in the target-droplet exceeds the error-tolerance limit in all
cases when a volumetric split-error occurs in the last but one step. Moreover, the {\em CF}-error in the target-{\em CF} increases when the 
magnitude of
volumetric split-error increases.

\subsection{Multiple volumetric split-errors}
\label{sec:multiple_error}
To this end, we have analyzed the effect of single volumetric split-error
on the target-{\em CF}~(with different erroneous-volume droplets) and observed that the {\em CF}-error in the
target-droplet increases when the magnitude of split-error increases. However, due
to unpredictable characteristics of fluid droplets, 
such split-errors may
occur in multiple mix-split steps of the mixing path; they may change
the {\em CF} of the desired target-droplet significantly. Moreover, 
volumetric split-errors may occur in any combination of signs (use of larger or
smaller droplet following a split step) on the mixing path 
during sample preparation. We derive expressions that capture the overall
effect of such errors on the target-{\em CF}.


{\color{black} Let $\epsilon_i$} indicate
the percentage of the volumetric split-error occurring at the
$i^{th}$ mix-split step.
A fundamental question in this context is the following:
\emph{``How is the {CF} of a target-droplet affected by multiple volumetric split-errors \{$\epsilon_1$, $\epsilon_2$,$\ldots$, $\epsilon_{i-1}$\} 
	occurring at different mix-split steps in the mixing path during sample preparation?''}. 

In order to find a reasonable answer to the above question, let us consider the dilution problem for generating a target-{\em CF} = $C_t$ 
using {\em twoWayMix}~{\cite{BS}} as shown in Fig.~\ref{fig:error}. Here, $O_i$ 
represents the $i^{th}$ (1:1) mix-split step, $C_i$ is the resulting 
{\em CF} after the $i^{th}$ mix-split step, and
$r_i$ is the {\em CF} of the source (100\% for sample, 0\% for buffer) used in $i^{th}$ mix-split operation.  Without loss of
generality, let us assume that a volumetric split-error $\epsilon_i$ occurs after the $i^{th}$ mix-split step of the mixing path, i.e., a two-unit volume 
droplet produces, after splitting, two daughter-droplets of volume 
1+$\epsilon$ and
1-$\epsilon$, $\epsilon>$0. Initially, sample  and
buffer are mixed at the first mix-split step ($O_1$). After this mixing operation, the {\em CF} and volume of the 
resulting droplet become 
{$C_1$ = $\frac{P_0 \times (1 \pm \epsilon_0) + 2^{-1} \times r_0}{Q_0 \times (1 \pm  \epsilon_0) + 2^{-1}}$}
and $V_1$ = $\frac{Q_0 \times (1 \pm \epsilon_0) + 2^{-1}}{2^0}$, respectively, where
{$P_0$ = $Q_0$ = $\frac{1}{2}$, $\epsilon_0$ = $r_0$ = 0}. Note that $r_i$ = 1 (0) indicates whether a sample (buffer) is used in the $i^{th}$ mix-split step of the mixing path. Furthermore, the sign $+$ ($-$) in the expression indicates whether a larger (smaller) droplet is used in the
next mix-split step followed by a split operation.

A volumetric split-error may occur in one or more mix-split operations of the mixing path while preparing a target-{\em CF}. For example, volumetric split-errors  \{$\epsilon_1$, $\epsilon_2$, $\ldots$, $\epsilon_{6}$\} may occur, one after another, in
	the mix-split operations \{$O_1$, $O_2$, $\ldots$, $O_6$\} as shown
	in Fig.~\ref{fig:error}. In Table~\ref{tab:equation}, we report the
	volume and concentration of the resulting daughter-droplets after each
	mix-split operation when all of the preceding steps suffer from split-errors.

\setlength{\tabcolsep}{2.6pt}
\begin{table*}[!h]
	\begin{center}
		\begin{threeparttable}
			\centering
			\caption{{Impact of split-errors on the resulting daughter-droplets.}}
			\scriptsize
			\label{tab:equation}
			\setlength{\extrarowheight}{.35em}
			\begin{tabular}{p{82pt}|p{63pt}|p{86pt}|p{74pt}|p{110pt}}
				\hline
				{{Erroneous mix-split step ($O_i$)}} & {Split-error}  &{{$\overline{CF}$}} & $\overline{V}$ & Parameter values\\
				 \hline
				 \{$O_1$\} & \{$\epsilon_1$\} & $C_2$ = $\frac{P_1 \times (1 \pm \epsilon_1) + r_1}{Q_1 \times (1 \pm  \epsilon_1) + 2^0}$ & {$V_2$ = $\frac{Q_1 \times (1 \pm \epsilon_1) + 2^0}{2}$} &
				 $P_1$ = $P_0 \times(1 \pm \epsilon_0) + 2^{-1} \times r_0$, {$Q_1$ = $Q_0\times(1 \pm  \epsilon_0) + 2^{-1}$} \\ \hline
				 
				 \{$O_1$, $O_2$\} &  \{$\epsilon_1$, $\epsilon_2$\} & $C_3$ = $\frac{P_2 \times (1 \pm \epsilon_2) + 2 \times r_2}{Q_2 \times (1 \pm \epsilon_2) + 2}$ & 
				 {$V_3$ = $\frac{Q_2 \times (1 \pm\epsilon_2) + 2}{2^2}$} & {$P_2$ = ${P_1\times(1 \pm \epsilon_1) + r_1}$},  $\hspace{14pt} Q_2$ = ${Q_1\times(1 \pm \epsilon_1) + 2^0}$ \\ \hline
				 
				 \{$O_1$, $O_2$, $O_3$\} &  \{$\epsilon_1$, $\epsilon_2$, $\epsilon_3$\} & $C_4$ = $\frac{P_3 \times (1 \pm \epsilon_3) + 2^2 \times r_3}{Q_3 \times (1 \pm \epsilon_3) + 2^2}$ & {$V_4$ = $\frac{Q_3 \times (1 \pm\epsilon_3) + 2^2}{2^3}$} & 
				 {$P_3$ = ${P_2\times(1\pm\epsilon_2)+2\times r_2}$}, $Q_3$ =  ${Q_2\times(1 \pm\epsilon_2) + 2}$ \\ \hline
				 $\ldots$ & $\ldots$ & $\ldots$ & $\ldots$ & $\ldots$ \\
				  $\ldots$ & $\ldots$ & $\ldots$ & $\ldots$ & $\ldots$ \\ \hline
				  
				 \{$O_1$,$O_2$,$O_3$,$O_4$,$O_5$,$O_6$\} &  \{$\epsilon_1$,$\epsilon_2$,$\epsilon_3$,$\epsilon_4$,$\epsilon_5$,$\epsilon_6$\} & $C_7$ = $\frac{P_6\times(1 \pm \epsilon_6) + 2^{5} \times r_6}{Q_6\times(1 \pm \epsilon_6) + 2^5}$ & $V_7$ = $\frac{Q_6 \times (1 \pm\epsilon_6) + 2^5}{2^6}$ & {$P_6$ = ${P_5\times(1 \pm \epsilon_5) + 2^4 \times r_5}$}, $Q_6$ =  ${Q_5\times(1 \pm\epsilon_5) + 2^4}$ \\
				 
				 & & \\
				  
				\hline
			\end{tabular}
	
			\begin{tablenotes}
				\item {{$\overline{CF}$}}: Concentration of the resulting daughter-droplets after next mix-split step; {{$\overline{V}$}}: Volume of the resulting daughter-droplets after next mix-split step.
			\end{tablenotes}
		\end{threeparttable}
	\end{center}
\end{table*}

\begin{figure*}[!b]
	\centering
	\includegraphics[width=10.8cm, height = 8.8cm]{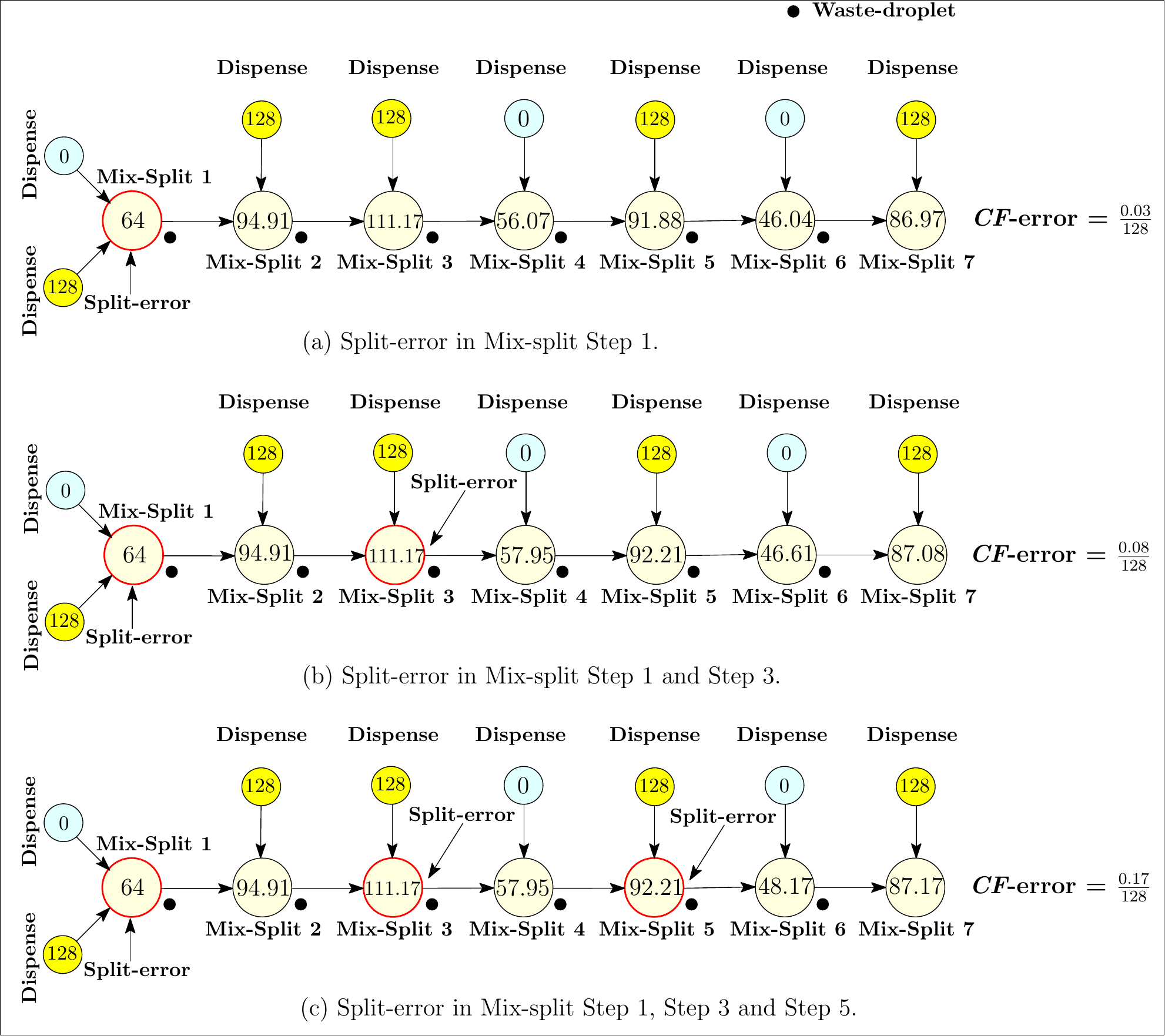}
	\caption{Effect of multiple volumetric split-errors on the target-{\em CF} = $\frac{87}{128}$.}
	\label{fig:multi_error_example_cf}
\end{figure*}

Hence, for the occurrence  of multiple volumetric split-errors, say  \{$\epsilon_1, \epsilon_2, \epsilon_3, \ldots, \epsilon_{i-2}, \epsilon_{i-1}$\} at mix-split steps \{$O_1$, $O_2$,, $O_3$, \ldots, $O_{i-2}$, $O_{i-1}$\}, the {\em CF} and volume of the generated target-droplet after
the final mix-split operation can be computed using the following expressions:
\begin{equation}
C_i = \frac{P_{i-1} \times (1 \pm \epsilon_{i-1}) + 2^{i-2} \times r_{i-1}}{Q_{i-1} \times (1 \pm \epsilon_{i-1}) + 2^{i-2}}
\label{lbl:con}
\end{equation} 
\begin{equation}
V_i = \frac{Q_{i-1} \times (1 \pm\epsilon_{i-1}) + 2^{i-2}}{2^{i-1}}
\label{lbl:vol}
\end{equation}

where $P_i$ = ${P_{i-1}\times (1 \pm \epsilon_{i-1}) + 2^{i-2} \times r_{i-1}}$ and {$Q_i$ =  ${Q_{i-1} \times (1 \pm\epsilon_{i-1}) + 2^{i-2}}$}.
In this way, the impact of multiple volumetric split-errors occurring on different mix-split steps of the mixing path on the  target-{\em CF} can be precomputed.

In order to find the effect of multiple volumetric split-errors on the target-{\em CF}, we
perform several experiments.
We continue with the example target-{\em CF} = $\frac{87}{128}$ of
accuracy level = 7, and inject 7\% volumetric split-error
simultaneously at different mix-split steps of the mixing path. 
The effects of such split-errors are shown in Fig.~\ref{fig:multi_error_example_cf}.
During simulation, we assume that the larger erroneous droplet is always used later when
a split-error occurs in the mix-split path~(i.e., $\epsilon$ is positive). For example, 
the effect of multiple 7\% volumetric split-errors in Mix-Split Step 1 and Step
3 is shown in Fig.~\ref{fig:multi_error_example_cf}(b). Only the effect of three
concurrent volumetric split-errors is also shown in Fig.~\ref{fig:multi_error_example_cf} (c). It has been observed that
{\em CF}-error in the target-droplet rapidly grows to $\frac{0.08}{128}$ and
$\frac{0.17}{128}$ when two or three such split-errors are injected in the mix-split path.

\begin{figure*}[!t]
	\centering
	\includegraphics[width=10.6cm, height = 9.0cm]{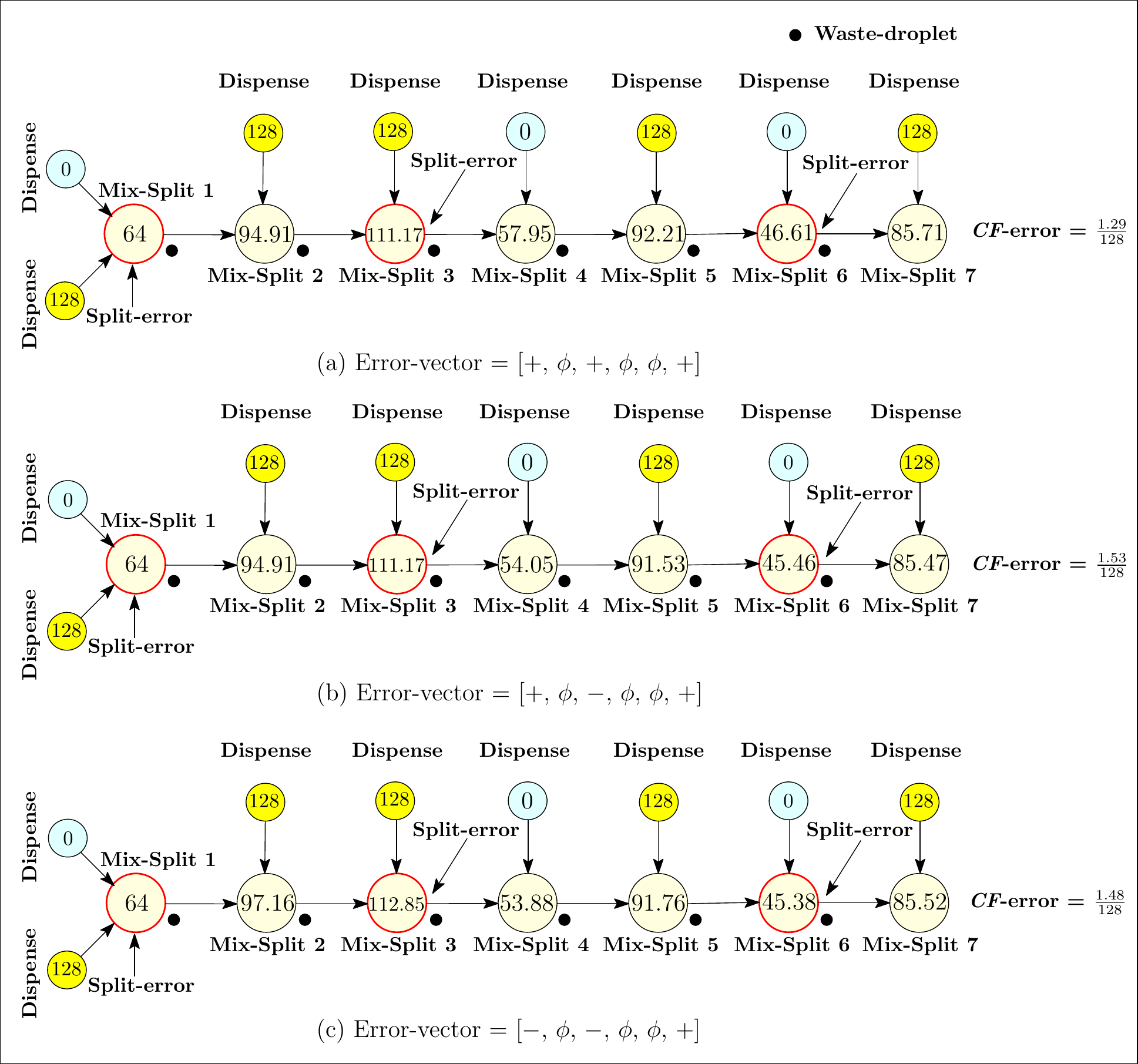}
	\caption{Effect of multiple volumetric split-errors on the target-{\em CF} = $\frac{87}{128}$.}
	\label{fig:multi_error_vector}
\end{figure*}

\section{Worst-case error in the target-$CF$}
\label{sec:max_error}
So far we have analyzed the effect of multiple volumetric split-errors
on a target-{\em CF} when a larger erroneous droplet is selected following each
mix-split step. However, in a ``sensor-free" environment, one cannot select 
the larger erroneous droplet at will for the
subsequent operations. In reality, multiple volumetric split-errors may 
consist of an arbitrary combination of large and small
daughter-droplets. Hence, further analysis is required to reveal the role
of such random occurrence of volumetric split-errors and their effects on
the target-{\em CF}.

In order to facilitate the analysis, we define ``error-vector" as follows:
An error-vector of length {\em k} denotes the sequence
of larger or smaller erroneous droplets, which are chosen corresponding to
{\em k} mix-split errors in the mixing path. For
example, an error-vector [+,$\phi$,$-$,$\phi$,$\phi$,+] denotes volumetric split-error in Mix-Split Step 1, Step 3, and Step 6, where $\phi$ denotes no-error. In Step 1, the larger droplet is passed to the next step, whereas 
in Step 3, the smaller one is used in the next step, and so on. For {\em k} volumetric split-errors,
$3^k$ error-vectors are possible. While executing actual mix-split
operations, the target-{\em CF} can be affected by any one of them.

We perform simulated experiments for finding the effect of different
error-vectors for the target-{\em CF} = $\frac{87}{128}$. Initially,
we observe the effect of three errors corresponding to the mix-split operations \{Mix-Split 1, Mix-Split 3, Mix-Split 6\}
to the target-{\em CF} (for 7\% split-error). See Fig.~\ref{fig:multi_error_vector} for an example.

We observe that {\em CF}-error in the target-droplet increases 
noticeably for the error-vectors [+,$\phi$,+,$\phi$,$\phi$,+], [+,$\phi$,$-$,$\phi$,$\phi$,+] and [$-$,$\phi$,$-$,$\phi$,$\phi$,+]
as depicted in the 
Fig.~\ref{fig:multi_error_vector} (a)-(c). It has been seen from the Fig.~\ref{fig:multi_error_vector} that {\em CF}-error exceeds the
error-tolerance limit~($\frac{0.5}{128}$) in each cases. Thus
target-{\em CF} is affected badly for these error-vectors. We also
perform similar experiments with volumetric split-error 3\% and found
that {\em CF}-error decreases for all cases.
  
Moreover, we perform simulation for revealing the effect
of remaining error-vectors on the target-{\em CF} and report the
generated {\em CF}s by all possible error-vectors (\# error-vectors = 8)
in Table.~\ref{tab:multi_error}. It has been observed that the
{\em CF}-error exceeds allowable error-tolerance limit in all such cases.
On the other hand, the maximum {\em CF}-error in the target-{\em CF} occurs
for the error-vector [$-$,$\phi$,+,$\phi$,$\phi$,$-$] which is $\frac{1.61}{128}$~($>$ error-tolerance limit).

\setlength{\tabcolsep}{.5pt}
\begin{table}[!t]
	\begin{center}
		\begin{threeparttable}
			\centering
			\caption{{Effect of different error-vectors on the target-{\em CF} = $\frac{87}{128}$ for split-error = 7\%.}}
			\scriptsize
			\label{tab:multi_error}
			\setlength{\extrarowheight}{.4em}
			\begin{tabular}{c|c|c|c}
				\hline
				{Error-vector} & {Produced {\em CF}$\times$128*}  &{Produced {\em CF}-error$\times$128} & {{\em CF}-error$\times$128 $<$ 0.5?}\\
				\hline
				[+, $\phi$, +, $\phi$, $\phi$, +] & 85.71 & 1.29 & No \\ \hline
				[+, $\phi$, +, $\phi$, $\phi$, $-$] & 88.56 & 1.56 & No \\ \hline
				[+,  $\phi$, $-$, $\phi$, $\phi$, +] & 85.47 & 1.53 & No \\ \hline
				[+, $\phi$, $-$, $\phi$, $\phi$, $-$] & 88.36 & 1.36 & No \\ \hline
				[$-$, $\phi$, +, $\phi$, $\phi$, +] & 85.76 & 1.24 & No \\ \hline
				[$-$, $\phi$, +, $\phi$, $\phi$, $-$] & 88.61 & 1.61 & No \\ \hline
				[$-$, $\phi$, $-$, $\phi$, $\phi$, +] & 85.52 & 1.48 & No \\ \hline
				[$-$, $\phi$, $-$, $\phi$, $\phi$, $-$] & 88.41 & 1.41 & No \\ \hline
			\end{tabular}
			
			\begin{tablenotes}
				\item *: Results are shown up to two decimal places.
			\end{tablenotes}
		\end{threeparttable}
	\end{center}
\end{table}

Note that volumetric split-error may also occur in the remaining mix-split
steps, i.e., each  mix-split step of the mixing path may suffer from
volumetric split-errors. Therefore, it is also essential to reveal the effect
of multiple volumetric split-errors on the target-{\em CF} when an error occurs in
each mix-split operation.

We further perform experiments to find the effect of such volumetric
split-errors on the target-{\em CF} = $\frac{87}{128}$. The mix-split
graph of the target-{\em CF} = $\frac{87}{128}$ consist of 7 mix-split operations~(see Fig.~\ref{fig:example_cf}). During simulation,
we inject split-error in each mix-split step of the mixing path except the
final mix-split operation~(Mix-Split Step 7) since any volumetric split-error
in the final mix-split operation will not alter the target-{\em CF}
anymore~(only the volumes of two resulting target-droplets may change).
So there will be
six potential mix-split steps~(except the final one) where split-error can occur. Thus, there will be 64
possible error-vectors. We set split-error = +0.07 or -0.07, in each
mix-split step, depending on the sign of the error in the corresponding 
position of vector. We perform experiments exhaustively and report the results for
some representative error-vectors for the target-{\em CF} = $\frac{87}{128}$ in Table~\ref{tab:full_error_error}. We see
that the {\em CF}-error exceeds the allowable error-range in every case.


\setlength{\tabcolsep}{.5pt}
\begin{table}[tp]
	\begin{center}
		\begin{threeparttable}
			\centering
			\caption{{Effect of some error-vectors of length 6 on the target-{\em CF} = $\frac{87}{128}$ for split-error =  7\%.}}
			\scriptsize
			\label{tab:full_error_error}
			\setlength{\extrarowheight}{.45em}
			\begin{tabular}{c|c|c|c}
				\hline
				{Error-vector} & {Produced {\em CF}$\times$128*}  &{Produced {\em CF}-error$\times$128} & {{\em CF}-error$\times$128 $<$ 0.5?}\\
				\hline
				[+,+,+,+,+,+] & 85.58 & 1.42 & No \\ \hline
				[+,$-$,+,+,+,+] & 85.53 & 1.47 & No \\ \hline
				[+,$-$,$-$,+,+,+] & 85.26 & 1.74 & No \\ \hline
				[+,$-$,+,+,$-$,+] & 85.08 & 1.92 & No \\ \hline
				[$-$,+,$-$,$-$,+,$-$] & 88.78 & 1.78 & No \\ \hline
				[$-$,+,+,$-$,$-$,$-$] & 88.82 & 1.82 & No \\ \hline
				[$-$,+,$-$,$-$,$-$,$-$] & 88.64 & 1.64 & No \\ \hline
				[$-$,$-$,$-$,$-$,$-$,$-$] & 88.61 & 1.61 & No \\ \hline
			\end{tabular}
			
			\begin{tablenotes}
				\item *: Results are shown up to two decimal places.
			\end{tablenotes}
		\end{threeparttable}
	\end{center}
\end{table}

\begin{figure}[t]
	\centering
	\includegraphics[width=9.5cm, height = 6.0cm]{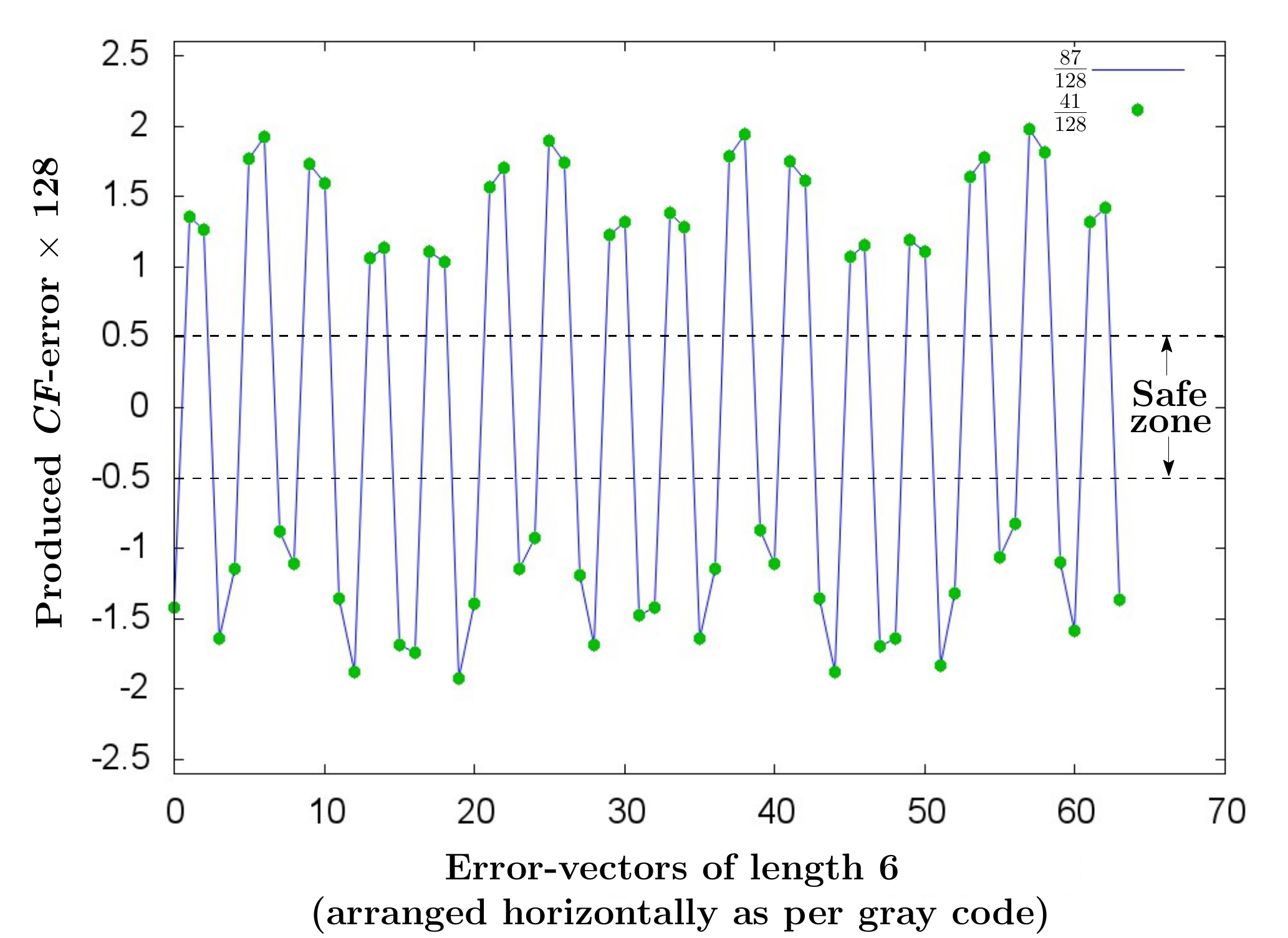}
	\caption{Value of ({\em CF}-error$\times$128) for all possible error-vectors with 7\% split-error for the target-{\em CF} = $\frac{41}{128}$ and $\frac{87}{128}$.}
	\label{fig:anal_87}
\end{figure}

\begin{figure}[!t]
	\centering
	\includegraphics[width=9.5cm, height = 6.0cm]{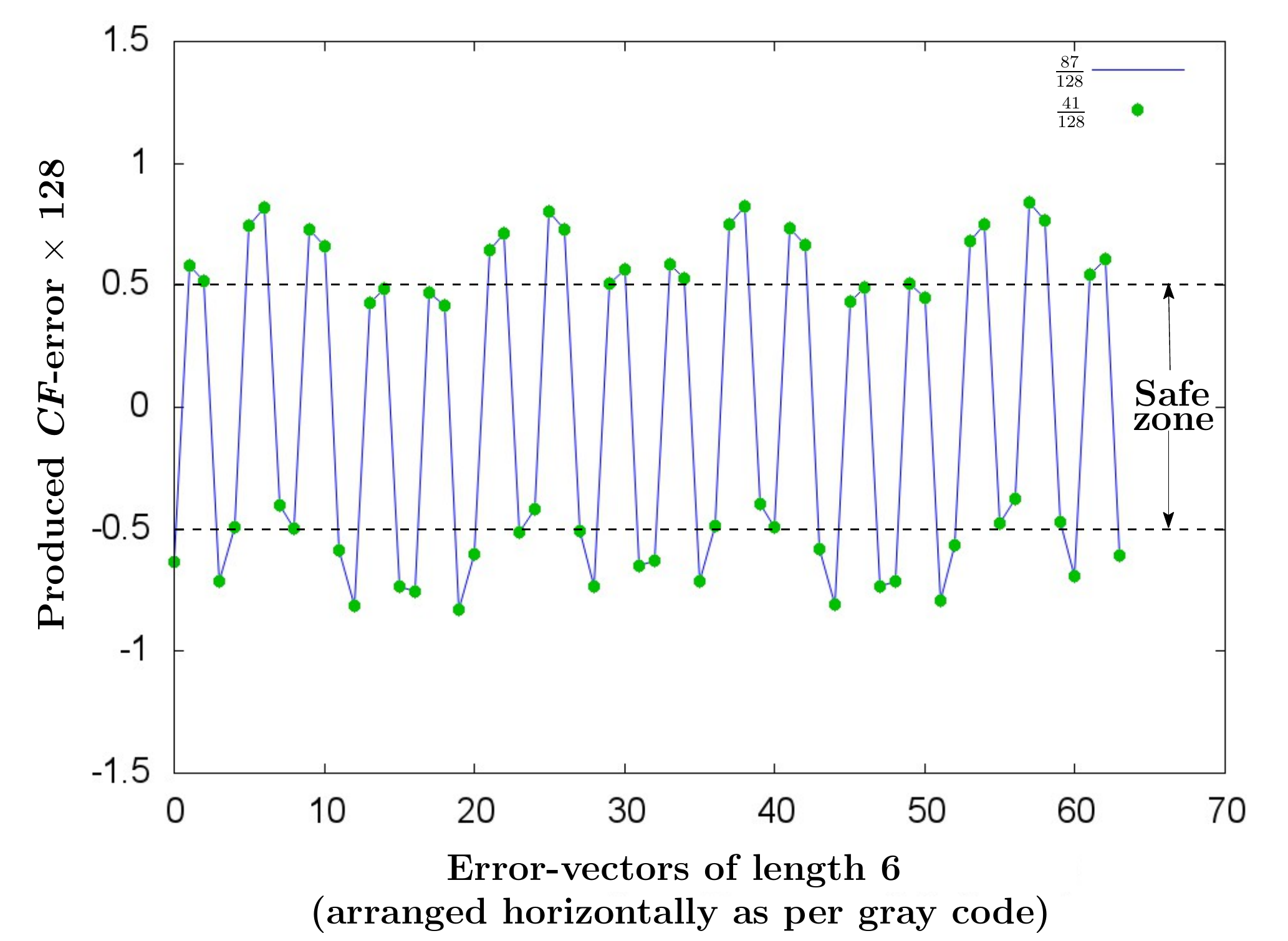}
	\caption{Value of ({\em CF}-error$\times$128) for all possible error-vectors with 3\% split-error for the target-{\em CF} = $\frac{41}{128}$ and $\frac{87}{128}$.}
	\label{fig:anal_87_3}
\end{figure}

\begin{figure}[!t]
	\centering
	\includegraphics[width=9.5cm, height = 6.0cm]{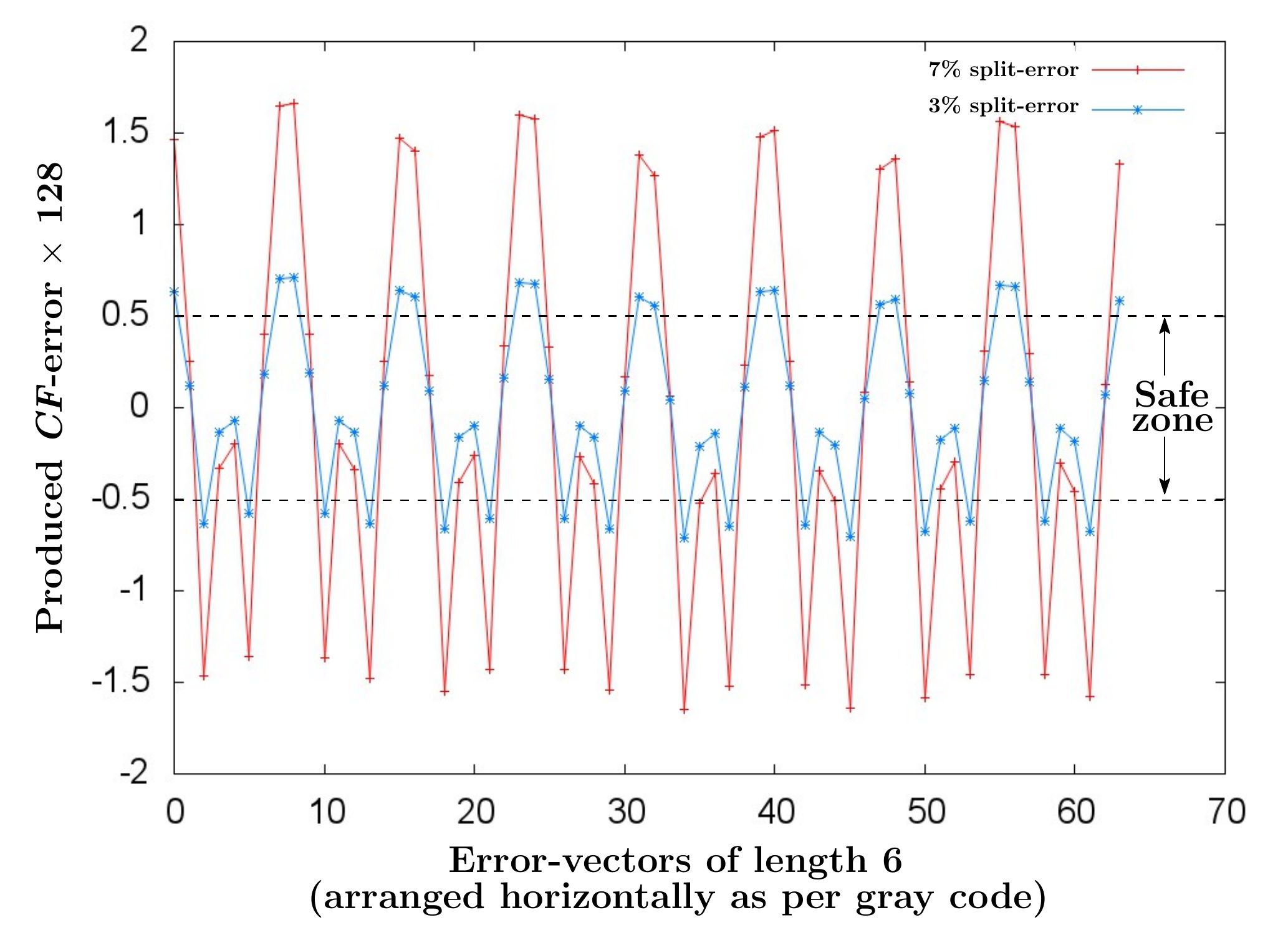}
	\caption{Value of ({\em CF}-error$\times$128) for all possible error-vectors for 7\% and 3\% split-error for the target-{\em CF} = $\frac{17}{128}$.}
	\label{fig:anal_17}
\end{figure}

We also 
show the {\em CF}-error by all possible error-vectors in
Fig.~\ref{fig:anal_87} for the target-{\em CF} = $\frac{41}{128}$ and 
$\frac{87}{128}$~(complement of $\frac{41}{128}$) for demonstration purpose.
We plot error-vectors (setting + $\rightarrow$ 0,
$-~\rightarrow$ 1) along the X-axis, and arrange them from left-to-right
following a gray-code, so that any two adjacent vectors are only unit
Hamming distance apart. The Y-axis shows the corresponding values of
{\em CF}-error$\times$128. Based on exhaustive simulation, we observe
that the {\em CF}-error in both target-{\em CF}s becomes maximum~(1.977)
for the error-vector [$-$,+,+,$-$,+,$-$] (at the 57$^{th}$ position on the
X-axis). Note that for the target-{\em CF} = $\frac{41}{128}$, 
{\em CF}-errors are multiplied with -1 for the ease of analysis. 
We notice that none of these outcomes lies within the safe-zone~(within error-tolerance limit).
We also perform similar experiment for both the target-{\em CF}s when
split-error becomes 3\% and observe
that for 12 cases, the errors lies within the tolerance zone, and the 
maximum {\em CF}-error reduces to $\frac{0.84}{128}$ corresponding to 
the same error-vector [$-$,+,+,$-$,+,$-$] (Fig.~\ref{fig:anal_87_3}) for
both {\em CF}s. However, for the target-{\em CF} = $\frac{17}{128}$, a large number 
of {\em CF}-errors = 29~(32) generated by all possible
error-vectors of length 6 lie within the error-tolerance zone for
7\% (3\%) split-errors (see Fig.~\ref{fig:anal_17}). Note that the 
magnitude of 
{\em CF}-errors decreases in each case when split-error
reduces to 3\%.

\begin{figure*}[!h]
	\centering
	\includegraphics[width=10.5cm, height = 6.0cm]{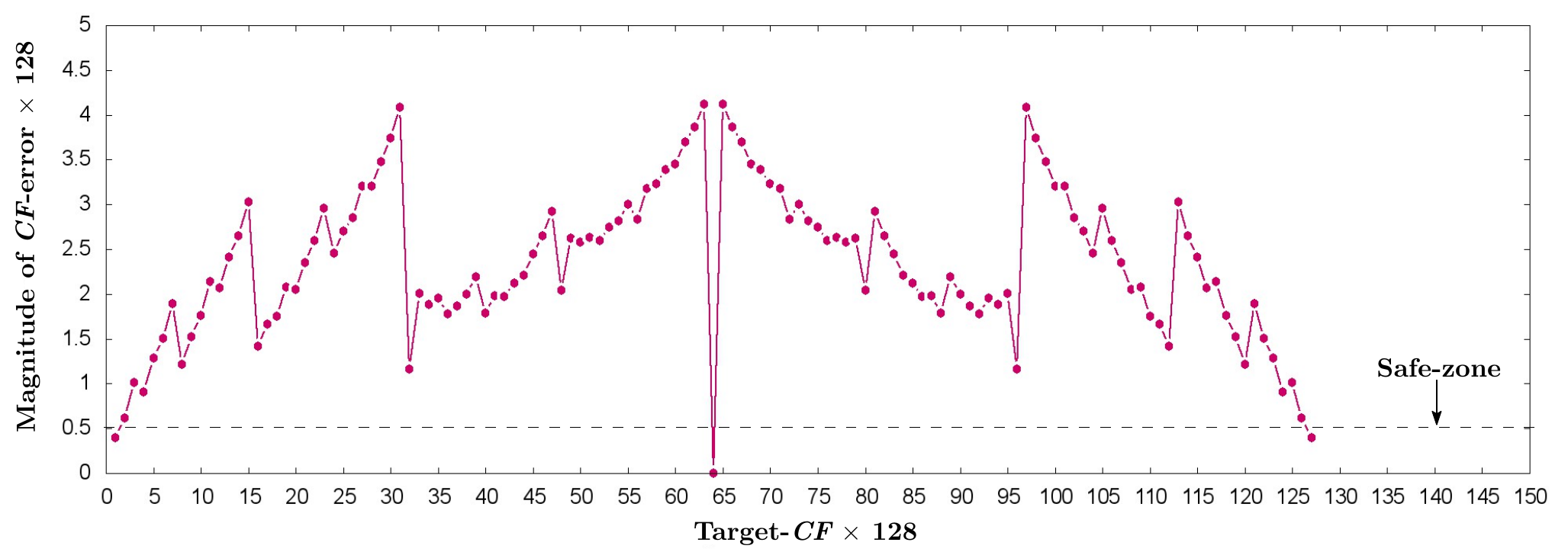}
	\caption{Maximum value of ({\em CF}-error$\times$128) for all target-{\em CF}s with accuracy level = 7.}
	\label{fig:anal_max_error}
\end{figure*}

We further perform simulation for measuring maximum {\em CF}-error
generated for all target-{\em CF}s of accuracy level 7~(with 7\% split-error). We plot the results in Fig.~\ref{fig:anal_max_error}.
We observe that the {\em CF}-error for the target-{\em CF} = $\frac{63}{128}$
and $\frac{65}{128}$ becomes maximum~($\frac{4.12}{128}$) compared to
those produced by other error-vectors. The error-vector [$-$, $-$, $-$, $-$, $-$, $-$] generates the maximum {\em CF}-error for both these
target-{\em CF}s.

\section{Maximum CF-error: A justification}
\label{sec:max_error_justifiaction}
Motivated by the need for a formal proof for generating maximum
{\em CF}-error in the target-{\em CF}, we have performed a rigorous
theoretical analysis and further experiments to study the properties of
{\em CF}-error in a target-{\em CF}. The following analysis, as
shown below, reveals how the problem of error-tolerance can be handled in a more concrete fashion.

Consider a particular target-{\em CF} = $C_t$ and its dilution tree.
Let
the current mix-split step be $i$ (other than the last step, where the occurrence of 
split-error does not matter), and the intermediate-{\em CF} arriving at $i$
be $C_i$. If a 1X sample (buffer) droplet is added in this step, it
produces {\em CF} = $\frac{C_i+1}{2}$  (= $\frac{C_i}{2}$), assuming that the volume of the droplet arriving at $i$ is correct (1X).

Consider the first case, and assume that the droplet arriving at $i$
suffers a volumetric split-error of magnitude $\epsilon$  at the previous
step. Hence, after mixing with a sample droplet, the intermediate-{\em CF} will become: $\frac{C_i(1+\epsilon)+1}{2+\epsilon}$; the sign of 
$\epsilon$ is set to positive (negative) when the incoming
intermediate-droplet is larger (smaller) than the ideal volume 1X.
Thus, the error ($E_r$) in the intermediate-{\em CF} becomes:

\begin{equation}
\label{Eq:anal_equ1}
E_r = \frac{C_i+1}{2} - \frac{C_i(1+\epsilon)+1}{2+\epsilon} = \frac{\epsilon(1-C_i)}{4+2\epsilon}
\end{equation}

\begin{figure}[t]
	\centering
	\includegraphics[width=9.5cm, height = 6.0cm]{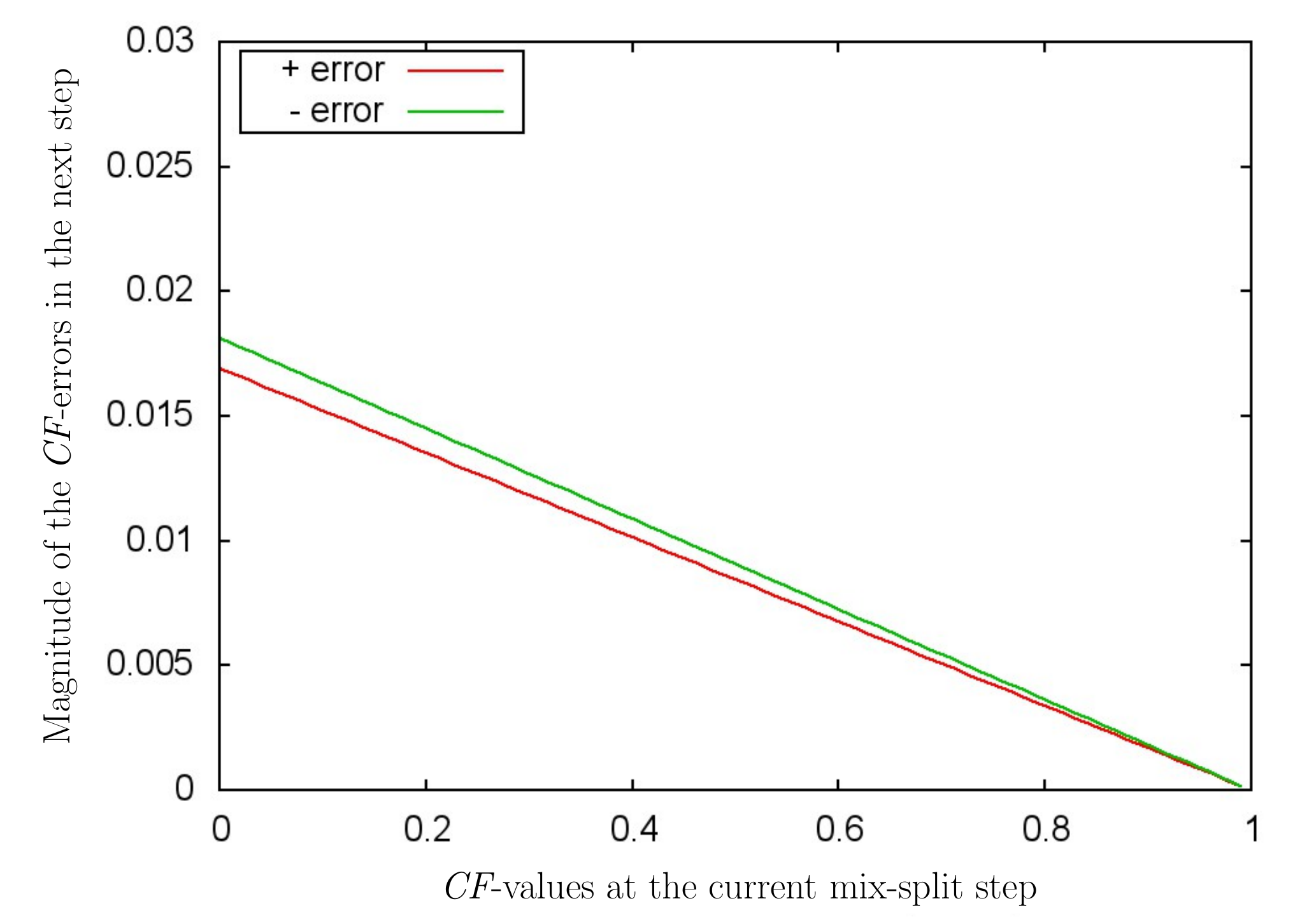}
	\caption{{\em CF}-error at the next mix-split step (for positive and negative single split-error).}
	\label{fig:anal_1}
\end{figure}

From Equation~\ref{Eq:anal_equ1}, it can be observed that the magnitude of
$E_r$ becomes larger when $\epsilon$ is negative, because a negative error
reduces the value of the denominator. In other words, the error in 
{\em CF} will be more if a droplet of smaller-volume arrives at Step $i$
compared to the case when a larger-volume droplet arrives at the mixer. In other
words, the {\em effect of the error is not symmetrical}; however, since the
volumes of the two daughters will be proportionately different as well,
when they are mixed, the error is canceled. We perform an experiment
assuming volumetric error (7\%), i.e., by setting 
$\epsilon$ = +0.07 or -0.07 in one mix-split step, for all values of
intermediate-{\em CF}s. The corresponding results are shown in Fig.~\ref{fig:anal_1}. It can be observed that a negative
split-error always produces larger {\em CF}-error in the target-{\em CF}
for a single split-error (error-vector of length 1). Similar effects will
be observed when a buffer droplet is mixed at Step $i$.

\begin{figure}[t]
	\centering
	\includegraphics[width=9.5cm, height = 6.0cm]{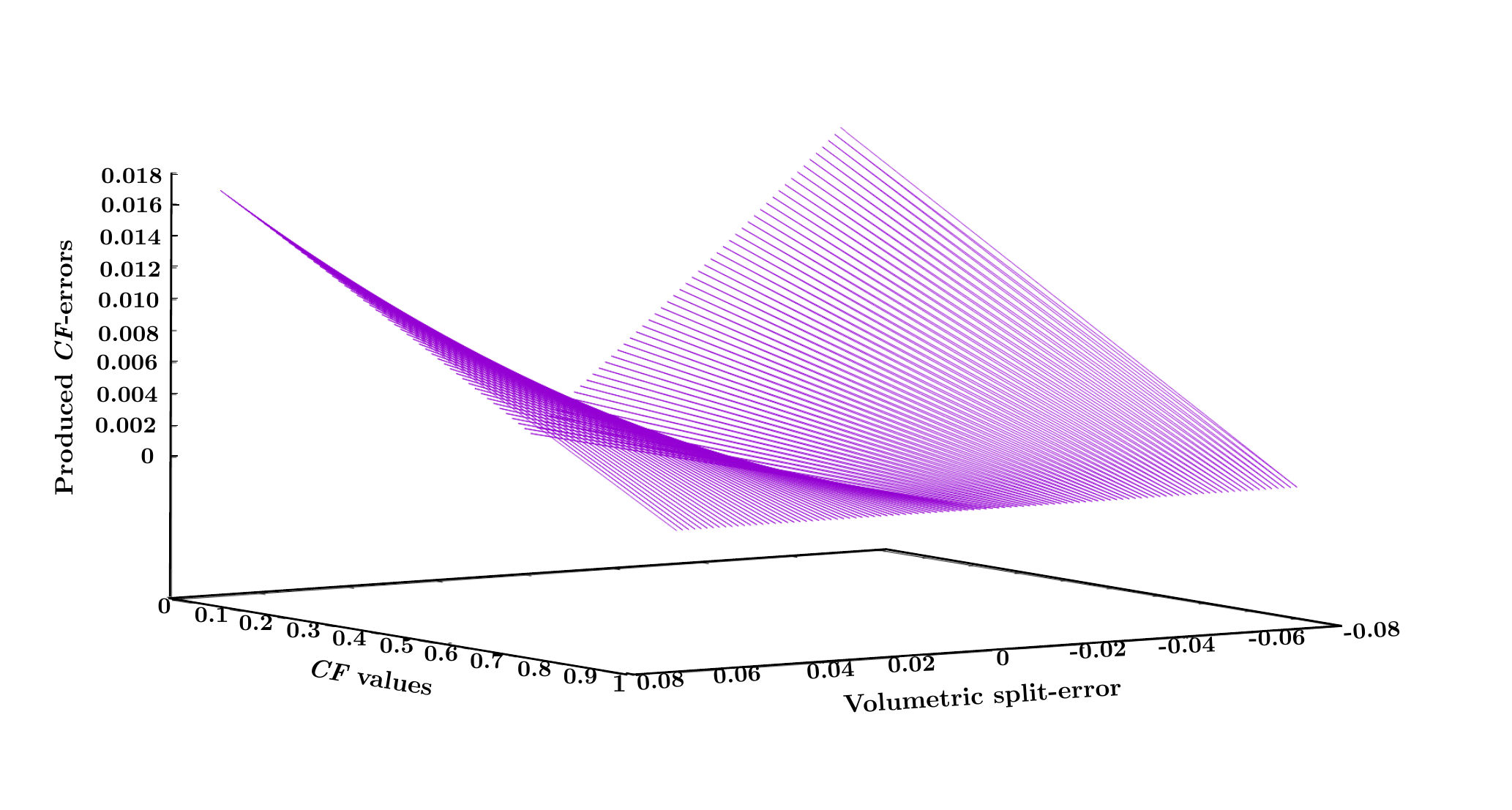}
	\caption{{\em CF}-error at the next mix-split step (for positive and negative single split-error).}
	\label{fig:anal_v1}
%
	\centering
	\includegraphics[width=9.5cm, height = 6.0cm]{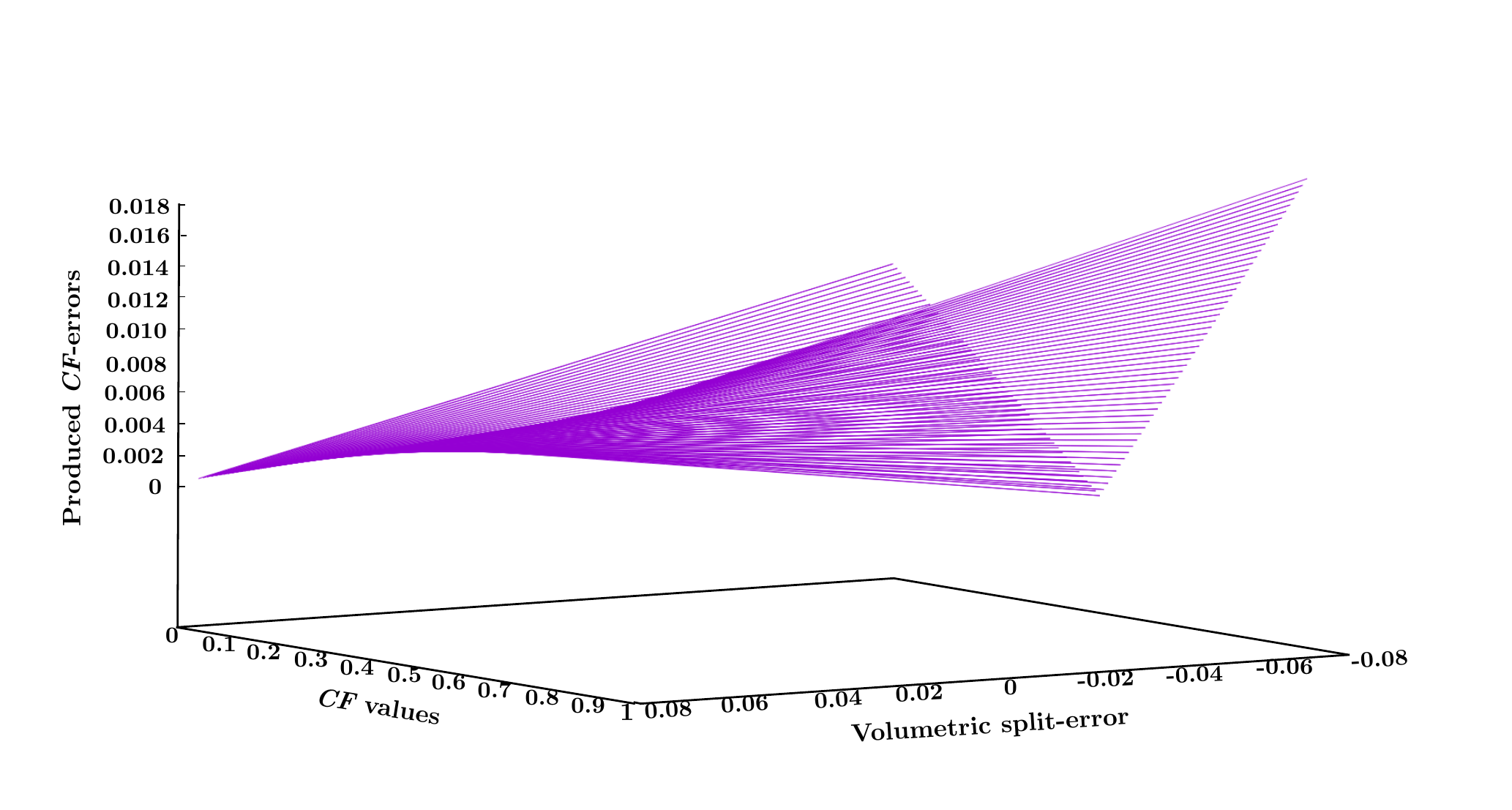}
	\caption{{\em CF}-error at the next mix-split step (for positive and negative single split-error).}
	\label{fig:anal_v2}
\end{figure}

We also perform simulation by varying $C_i$ from 0 to 1, and 
$\epsilon$ from -0.07 to 0.07 in Equation~\ref{Eq:anal_equ1} and calculate {\em CF}-errors.
We report the results as 3-dimensional~(3D) plots~(with different views) in
Fig.~\ref{fig:anal_v1} and Fig.~\ref{fig:anal_v2}, respectively.
We observe that simulation results favorably match with 
theoretical results~(see Fig.~\ref{fig:anal_1}), i.e., the negative
split-error~(single) always produces larger {\em CF}-error for a single
split-error. However, the effect
of error on a target-{\em CF} becomes much more complicated when multiple
split-errors are considered.

In order to demonstrate the intricacies, we have performed a representative analysis considering three consecutive split-errors. For simplicity, let us assume that an error of magnitude $\epsilon$ is injected in each of these three mix-split steps. Generalizing Equation~\ref{Eq:anal_equ1}, we can show that the corresponding {\em CF}-error observed after three steps will be:

\begin{multline*}
\footnotesize
\label{Eq:anal_equ2}
E_r = \frac{((((C_i(1+\epsilon)+r_1)(1+\epsilon)+r_2))(1+\epsilon)+r_3)}{(((2+(2+\epsilon)(1+\epsilon)))(1+\epsilon)+4)} - \\ \frac{(C_i+r_1+r_2+r_3)}{8} 
\end{multline*}

~~~~~~~~where $r_i$ = 1 (for sample droplet)

\hspace*{14mm} = 0 (for buffer droplet), for Step $i$, $i$ = 1, 2, 3.

\begin{figure}[!h]
	\centering
	\includegraphics[width=9.5cm, height = 6.0cm]{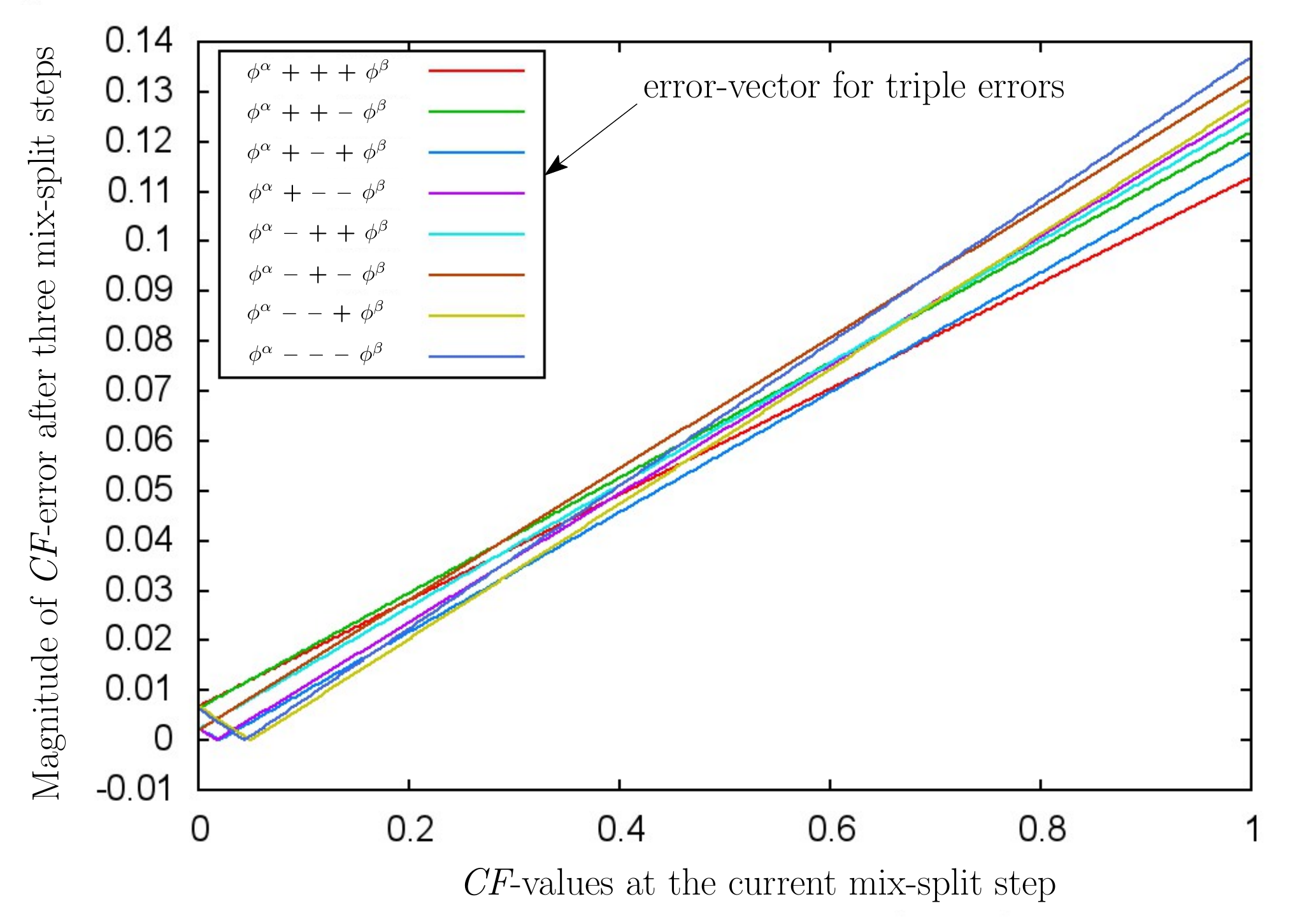}
	\caption{{\em CF}-error for triple split-errors.}
	\label{fig:anal_2}
\end{figure}

As before, we assume that $\epsilon$  = +0.07 or -0.07, and since we have
three consecutive split steps, we can have eight possible combinations of
such error vectors [$\phi^{\alpha}$, $-$, $-$, $-$, $\phi^{\beta}$], [$\phi^{\alpha}$, $-$, $-$, +, $\phi^{\beta}$], $\ldots$ , [$\phi^{\alpha}$, +, +, +, $\phi^{\beta}$] for a given 
combination of {$r_1$, $r_2$, $r_3$}, where $0\leq \alpha \leq$ ($n$ - 4), $0\leq \beta \leq$ ($n$ - 4) and $\alpha + \beta +$ 3  = $n$ - 1 ($n$ is the accuracy level). Thus, altogether, there will be 8
combinations. Fig.~\ref{fig:anal_2} shows the errors in {\em CF} observed
after three consecutive split-errors by setting {$r_1$ = 0, $r_2$ = 1,
$r_3$ = 1}, for all values of starting-{\em CF}, and for all eight
combinations of error-vectors. From the nature of the plot, it is apparent
that it is very hard to predict for which error-vector the maximum 
{\em CF}-error will occur, even for a given combination of $r$-values. The
maximum error depends on the {\em CF}-value from which the
{\em critical}-split-section begins and also on the error-vector that is chosen
(i.e., whether to proceed with the larger or the smaller
daughter-droplet). Furthermore, the error-expression becomes increasingly 
complex when the number of split-errors becomes large. 

As an example, we perform experiments to study the fluctuations of
the error in a
particular target-{\em CF} for all combinations of error-vectors and
showed the plot in Fig.~\ref{fig:anal_87}. Note that there are several peaks up and down in~\ref{fig:anal_87}, and based on exhaustive simulation, the value of (maximum-error $\times$ 128) is observed to be 1.977, which occurs for the error vector [$-$, +, +, $-$, +, $-$] (at the 57$^{th}$ position on the X-axis in Fig.~\ref{fig:anal_87}).

From the above analysis and experimental results, we conclude that it is 
hard to formulate a mechanism that will identify the exact
``maximum-error-vector" without doing exhaustive simulation. In other
words, it may not be possible to develop a procedure that will generate 
the maximum-error-vector without doing exhaustive analysis.

\section{Conclusion}
\label{sec:conclusion}
In this paper, initially, we have analyzed the effect of single volumetric 
split-errors (using the larger- or smaller-volume erroneous daughter-droplet) and found both theoretically and experimentally that the {\em CF}-error in
the target-droplet becomes larger when the smaller-volume daughter droplet is used in the assay, i.e., when $\epsilon$ becomes negative.
%
We also observed that the {\em CF}-error in a target-droplet increases with increasing magnitude of the split-error.  Next, we have performed various experiments to observe the
effect of multiple {\em CF}-errors on the target-{\em CF} and noticed that
it may be affected by any combination of erroneous droplets~(smaller/larger) during the execution of mix-split operations. We also observed that the {\em CF}-error in a
target-droplet increases when the target-{\em CF} is affected by a large
number
of split-errors. We performed rigorous analysis to identify the 
error vector that causes the maximum {\em CF}-error in the target-droplet. Unfortunately, it appears that it is very difficult to come up with an
algorithmic solution for identifying an error vector that maximizes the 
{\em CF}-error in the target under multiple split-errors.
But still, the observations and findings summarized in this paper will provide useful inputs to the development of methods for sample preparation that can deal with split errors even without any sensors and/or rollback (such as recently presented e.g.~in~\cite{error_oblivious,robust_sample_prep_2019}).

\ifCLASSOPTIONcaptionsoff
  \newpage
\fi



\bibliographystyle{IEEEtran}
\bibliography{ref_bib}

\begin{thebibliography}{10}
\providecommand{\url}[1]{#1}
\csname url@samestyle\endcsname
\providecommand{\newblock}{\relax}
\providecommand{\bibinfo}[2]{#2}
\providecommand{\BIBentrySTDinterwordspacing}{\spaceskip=0pt\relax}
\providecommand{\BIBentryALTinterwordstretchfactor}{4}
\providecommand{\BIBentryALTinterwordspacing}{\spaceskip=\fontdimen2\font plus
\BIBentryALTinterwordstretchfactor\fontdimen3\font minus
  \fontdimen4\font\relax}
\providecommand{\BIBforeignlanguage}[2]{{%
\expandafter\ifx\csname l@#1\endcsname\relax
\typeout{** WARNING: IEEEtran.bst: No hyphenation pattern has been}%
\typeout{** loaded for the language `#1'. Using the pattern for}%
\typeout{** the default language instead.}%
\else
\language=\csname l@#1\endcsname
\fi
#2}}
\providecommand{\BIBdecl}{\relax}
\BIBdecl

\bibitem{EWOD}
F.~Mugele and J.-C. Baret, ``Electrowetting: from basics to applications,''
  \emph{Journal of Physics: Condensed Matter}, vol.~17, no.~28, pp. 705--774,
  2005.

\bibitem{b19}
K.~Chakrabarty and F.~Su, \emph{{Digital Microfluidic Biochips - Synthesis,
  Testing, and Reconfiguration Techniques}}.\hskip 1em plus 0.5em minus
  0.4em\relax CRC Press, 2007.

\bibitem{B403341H}
V.~Srinivasan, V.~K. Pamula, and R.~B. Fair, ``An integrated digital
  microfluidic lab-on-a-chip for clinical diagnostics on human physiological
  fluids,'' \emph{Lab Chip}, vol.~4, pp. 310--315, 2004.

\bibitem{5487469}
K.~Chakrabarty, R.~B. Fair, and J.~Zeng, ``{Design tools for digital
  microfluidic biochips: Toward functional diversification and more than
  Moore},'' \emph{IEEE Trans. on CAD}, vol.~29, no.~7, pp. 1001--1017, 2010.

\bibitem{Alistar:2015}
M.~Alistar, P.~Pop, and J.~Madsen, ``{Redundancy Optimization for Error
  Recovery in Digital Microfluidic Biochips},'' \emph{Design Automation for
  Embedded Systems}, vol.~19, no. 1-2, pp. 129--159, 2015.

\bibitem{BS}
W.~Thies, J.~P. Urbanski, T.~Thorsen, and S.~P. Amarasinghe, ``Abstraction
  layers for scalable microfluidic biocomputing,'' \emph{Natural Computing},
  vol.~7, no.~2, pp. 255--275, 2008.

\bibitem{Dynamic}
Y.-L. Hsieh, T.-Y. Ho, and K.~Chakrabarty, ``{Biochip Synthesis and Dynamic
  Error Recovery for Sample Preparation Using Digital Microfluidics},''
  \emph{IEEE Trans. on CAD}, vol.~33, no.~2, pp. 183--196, 2014.

\bibitem{Poddar_2016}
S.~Poddar, S.~Ghoshal, K.~Chakrabarty, and B.~B. Bhattacharya,
  ``Error-correcting sample preparation with cyberphysical digital microfluidic
  lab-on-chip,'' \emph{ACM TODAES}, vol.~22, no.~1, pp. 2:1--2:29, 2016.

\bibitem{DMRW}
S.~Roy, B.~B. Bhattacharya, and K.~Chakrabarty, ``Optimization of dilution and
  mixing of biochemical samples using digital microfluidic biochips,''
  \emph{IEEE Trans. on CAD}, vol.~29, pp. 1696--1708, 2010.

\bibitem{REMIA}
J.-D. Huang, C.-H. Liu, and T.-W. Chiang, ``Reactant minimization during sample
  preparation on digital microfluidic biochips using skewed mixing trees,'' in
  \emph{Proc. of ICCAD}, 2012, pp. 377--384.

\bibitem{WARA}
C.-H. Liu, T.-W. Chiang, and J.-D. Huang, ``{Reactant Minimization in Sample
  Preparation on Digital Microfluidic Biochips},'' \emph{IEEE Trans. on CAD},
  vol.~34, no.~9, pp. 1429--1440, 2015.

\bibitem{MTC}
D.~Mitra, S.~Roy, S.~Bhattacharjee, K.~Chakrabarty, and B.~B. Bhattacharya,
  ``{On-Chip Sample Preparation for Multiple Targets Using Digital
  Microfluidics},'' \emph{IEEE Trans. on CAD}, vol.~33, no.~8, pp. 1131--1144,
  2014.

\bibitem{FloSPA}
S.~Bhattacharjee, S.~Poddar, S.~Roy, J.-D. Huang, and B.~B. Bhattacharya,
  ``Dilution and mixing algorithms for flow-based microfluidic biochips,''
  \emph{IEEE Trans. on CAD}, vol.~36, no.~4, pp. 614--627, 2017.

\bibitem{RSM}
Y.-L. Hsieh, T.-Y. Ho, and K.~Chakrabarty, ``{A Reagent-Saving Mixing Algorithm
  for Preparing Multiple-Target Biochemical Samples Using Digital
  Microfluidics},'' \emph{IEEE Trans. on CAD}, vol.~31, no.~11, pp. 1656--1669,
  2012.

\bibitem{stor_aware_2018}
S.~Bhattacharjee, R.~Wille, J.-D. Huang, and B.~Bhattacharya, ``Storage-aware
  sample preparation using flow-based microfluidic lab-on-chip,'' in
  \emph{Proc. of DATE}, 2018, pp. 1399--1404.

\bibitem{error_oblivious}
S.~Poddar, R.~Wille, H.~Rahaman, and B.~B. Bhattacharya, ``{Error-Oblivious
  Sample Preparation with Digital Microfluidic Lab-on-Chip},'' \emph{IEEE
  Trans. on CAD}, 2018, doi: \url{{10.1109/TCAD.2018.2864263}}.

\bibitem{control_path}
Y.~Zhao, T.~Xu, and K.~Chakrabarty, ``Integrated control-path design and error
  recovery in the synthesis of digital microfluidic lab-on-chip.'' \emph{{ACM
  JETC}}, vol.~6, no.~3, pp. 11:1--11:28, 2010.

\bibitem{cyber_physical}
Y.~Luo, K.~Chakrabarty, and T.-Y. Ho, ``{Error Recovery in Cyberphysical
  Digital Microfluidic Biochips},'' \emph{IEEE Trans. on CAD}, vol.~32, no.~1,
  pp. 59--72, 2013.

\bibitem{Luo_dictionary}
------, ``Real-time error recovery in cyberphysical digital-microfluidic
  biochips using a compact dictionary,'' \emph{{IEEE} Trans. on {CAD}},
  vol.~32, no.~12, pp. 1839--1852, 2013.

\bibitem{LuoTCAD14Uncertainity}
------, ``{Biochemistry Synthesis on a Cyberphysical Digital Microfluidics
  Platform Under Completion-Time Uncertainties in Fluidic Operations},''
  \emph{IEEE Trans. on CAD}, vol.~33, no.~6, pp. 903--916, 2014.

\bibitem{Mein2000}
C.~A. Mein, B.~J. Barratt, M.~G. Dunn, T.~Siegmund, A.~N. Smith, L.~Esposito,
  S.~Nutland, H.~E. Stevens, A.~J. Wilson, M.~S. Phillips, N.~Jarvis, S.~Law,
  M.~D. Arruda, and J.~A. Todd, ``{Evaluation of single nucleotide polymorphism
  typing with invader on pcr amplicons and its automation.}'' \emph{Genome
  Research}, vol.~10, no.~3, pp. 330--343, 2000.

\bibitem{robust_sample_prep_2019}
Z.~Zhong, R.~Wille, and K.~Chakrabarty, ``Robust sample preparation on low-cost
  digital microfluidic biochips,'' in \emph{Proc. of {ASP-DAC}}, 2019.

\end{thebibliography}
\end{document}